\documentclass[useAMS]{mn2e}
\usepackage{latexsym,graphicx}

\usepackage{color}

\newcommand\kms{{\rm\,km\,s^{-1}}}
\newcommand\msun{\rm\,M_\odot}

\newcommand\hii{H\,{\sc ii} \,}

\newcommand{\MC}{\multicolumn}

\def\apgt{\ {\raise-.5ex\hbox{$\buildrel>\over\sim$}}\ }
\def\aplt{\ {\raise-.5ex\hbox{$\buildrel<\over\sim$}}\ }

\title[MCSNR\,J0127$-$7332 and SXP\,1062]{SALT observations of the supernova remnant 
MCSNR\,J0127$-$7332 and its associated Be X-ray binary SXP\,1062 in the SMC}
\author[V. V.~Gvaramadze et al.]
       {V. V.~Gvaramadze,$^{1,2,3}$\thanks{E-mail: vgvaram@mx.iki.rssi.ru}
     A. Y.~Kniazev,$^{4,5,1}$ J. S.~Gallagher,$^6$ L. M.~Oskinova,$^{7,8}$ \\
     \newauthor Y.-H.~Chu,$^{9}$ R. A.~Gruendl$^{10,11}$ and I. Y. Katkov$^{12,13,1}$ \\
    $^{1}$Sternberg Astronomical Institute, Lomonosov Moscow State University, Universitetskij Pr. 
    13, Moscow 119992, Russia\\
    $^{2}$Space Research Institute, Russian Academy of Sciences, Profsoyuznaya 84/32, Moscow 117997,
    Russia \\
    $^{3}$E. Kharadze Georgian National Astrophysical Observatory, Abastumani 0301, Georgia \\
    $^{4}$South African Astronomical Observatory, PO Box 9, 7935 Observatory, Cape Town, South Africa \\
    $^{5}$Southern African Large Telescope Foundation, PO Box 9, 7935 Observatory, Cape Town, South 
    Africa \\
    $^{6}$Department of Astronomy, University of Wisconsin-Madison, 5534 Sterling, 475 North Charter 
    Street, Madison, WI 53706, USA \\
    $^{7}$Institute for Physics and Astronomy, University Potsdam, 14476 Potsdam, Germany \\
    $^{8}$Kazan Federal University, Kremlevskaya Str 18, Kazan, Russia \\
    $^{9}$Institute of Astronomy and Astrophysics, Academia Sinica (ASIAA), 10617 Taipei, Taiwan \\
    $^{10}$Department of Astronomy, University of Illinois at Urbana-Champaign, 1002 W. Green Street, 
    Urbana, IL 61801, USA \\
    $^{11}$National Center for Supercomputing Applications, 1205 West Clark St., Urbana, IL 61801, USA \\
    $^{12}$New York University Abu Dhabi, P.O. Box 129188, Abu Dhabi, UAE \\
    $^{13}$Center for Astro, Particle, and Planetary Physics, NYU Abu Dhabi, PO Box 129188, Abu Dhabi, UAE \\
    }
\begin{document}

\date{Accepted 2021 March 04. Received 2021 February 18; in original form 2020 August 01}


\maketitle

\label{firstpage}

\begin{abstract}
We report the results of optical spectroscopy of the Small Magellanic Cloud supernova 
remnant (SNR) MCSNR\,J0127$-$7332 and the mass donor Be star, 2dFS\,3831, in its associated high-mass 
X-ray binary SXP\,1062 carried out with the Southern African Large Telescope (SALT). Using 
high-resolution long-slit spectra, we measured the expansion velocity of the SNR shell of $\approx140 
\, \kms$, indicating that MCSNR\,J0127$-$7332 is in the radiative phase. We found that the observed 
line ratios in the SNR spectrum can be understood if the local interstellar medium is ionized by 
2dFS\,3831 and/or OB stars around the SNR. We propose that MCSNR\,J0127$-$7332 is the result of 
supernova explosion within a bubble produced by the stellar wind of the supernova progenitor and that 
the bubble was surrounded by a massive shell at the moment of supernova explosion. We estimated the 
age of MCSNR\,J0127$-$7332 to be $\la10\,000$\,yr. We found that the spectrum of 2dFS\,3831 changes 
with orbital phase. Namely, the equivalent width of the H\,$\alpha$ emission line decreased by 
$\approx40$ per cent in $\approx130$\,d after periastron passage of the neutron star and then almost 
returned to its original value in the next $\approx100$\,d. Also, the spectrum of 2dFS\,3831 obtained 
closest to the periastron epoch (about three weeks after the periastron) shows a noticeable emission 
line of He\,{\sc ii} $\lambda$4686, which disappeared in the next about two weeks. We interpret these 
changes as a result of the temporary perturbation and heating of the disk as the neutron star passes 
through it.
\end{abstract}


\section{Introduction}
\label{sec:int}

A binary system surviving a supernova (SN) explosion of one of its components could evolve into 
an X-ray binary in which the compact stellar remnant (neutron star or black hole) accretes 
material from the normal (massive or low-mass) star. The typical time-scale for the formation of 
X-ray binaries containing neutron stars of $\sim10^6-10^8$\,yr (e.g. Tauris \& van den Heuvel 2006) 
is one to three orders of magnitude longer than the time-scale for visibility of supernova remnants 
(SNRs) of $\sim10^5$\,yr (e.g. Lozinskaya 1992), suggesting that none of such X-ray binaries should 
be detected within SNRs. However, several neutron star X-ray binaries (NSXRBs) were found to be 
associated with SNRs, which challenges the traditional view on the formation time-scales for these 
objects and is still waiting to be explained.

Of the known NSXRB/SNR associations, only one was found in our Galaxy, namely SNR\,G322.1+00.0/Cir\,X-1 
(Heinz et al. 2013; Linares et al. 2010). Two others were detected in the Small Magellanic Cloud (SMC): 
MCSNR\,J0127$-$7332/SXP\,1062 (H\'enault-Brunet et al. 2012; Haberl et al. 2012) and 
MCSNR\,J0103$-$7201/SXP\,1323 (Gvaramadze, Kniazev \& Oskinova 2019). And two more were found in the 
Large Magellanic Cloud: MCSNR\,J0536$-$6735/CXOU\,J053600.0$-$673507 (Seward et al. 2012; Corbet 
et al. 2016; van Soelen et al. 2019) and MCSNR\,J0513$-$6724/XMMU\,J051342.6$-$672412 (Maitra et al. 
2019). Study of these and similar systems could provide useful information on the magnetic and spin 
properties of young neutron stars, supernova kick velocities, parameters of pre-SN binaries, and 
evolution of post-SN orbits in NSXRBs (see, e.g., Haberl et al. 2012; Gonz\'alez-Gal\'an et al. 2018; 
Wang \& Tong 2020; Ho et al. 2020).

In this paper, we report the results of observations of the SNR MCSNR\,J0127$-$7332 in the 
wing of the SMC and the mass donor star, 2dFS\,3831, in its associated NSXRB SXP\,1062 
with the Southern African Large Telescope (SALT). In Section\,\ref{sec:smc}, 
we briefly review what is already known about these objects. Section\,\ref{sec:obs} describes our 
observations and data reduction. The obtained results are presented in Section\,\ref{sec:res} and 
discussed in Section\,\ref{sec:dis}. We summarise in Section\,\ref{sec:sum}.

\section{MCSNR\,J0127$-$7332/SXP\,1062: observational data}
\label{sec:smc}

The NSXRB SXP\,1062 was discovered by H\'enault-Brunet et al. (2012) in the wing of the SMC in the 
course of observations of the massive star-forming region NGC\,602 with {\it Chandra} and {\it XMM-Newton}. 
Like most of NSXRBs in the SMC, SXP\,1062 belongs to a class of Be X-ray binaries (BeXBs) that consist of 
a neutron star accreting from the circumstellar disc of a Be star. The neutron star in SXP\,1062 orbits 
around the B0.5(III)e star 2dFS\,3831 (H\'enault-Brunet et al. 2012) with an orbital period of 
$P_{\rm orb}\approx656$\,d (Schmidtke, Cowley \& Udalski 2012, 2019; see also 
Section\,\ref{sec:star}) and its spin period of $P_{\rm spin}=1062$\,s makes SXP\,1062 the third 
longest-period BeXB in the SMC (Haberl \& Sturm 2016). 

H\'enault-Brunet et al. (2012) presented spectra of 2dFS\,3831 obtained with the VLT-FLAMES
instrument on 2010 October 25 and the multi-fibre 2-degree Field (2dF) instrument of the Anglo-Australian 
Telescope in 1998 September (blue spectrum) and 1999 September (red spectrum). These spectra showed that 
the H\,$\alpha$ and H\,$\beta$ lines are purely in emission and revealed the presence of emission in the 
cores of other Balmer lines and apparent infilling of the He\,{\sc i} absorption lines. Also, emission 
lines of Fe\,{\sc ii} $\lambda$4179 and Fe\,{\sc ii} $\lambda$4233 and a weak absorption line of 
He\,{\sc ii} $\lambda$4542 were detected in the VLT-FLAMES spectrum, and a hint of weak He\,{\sc ii} 
$\lambda$4696 line was found in the 2dF spectrum. Gonz\'alez-Gal\'an et al. (2018) compiled equivalent 
width (EW) measurements for the H\,$\alpha$ line in the spectra of 2dFS\,3831 obtained in 1999--2016 and
found that its absolute value increased from $23\pm1$ \AA \, in 1999\footnote{Note that Gonz\'alez-Gal\'an 
et al. (2018) mistakenly indicated the date of this measurement as 2010 October 25, while actually it was 
obtained in 1999 September.} to $36.3\pm0.6$ \AA \, in 2014 and then decreased to $33\pm1$ \AA \, in the 
next two years after the X-ray outburst in mid-2014 (see also Section\,\ref{sec:star}).

H\'enault-Brunet et al. (2012) also discovered a SNR around SXP\,1062 and presented its H\,$\alpha$, 
[S\,{\sc ii}] and [O\,{\sc iii}] images from the Magellanic Cloud Emission-line Survey (MCELS),
and the higher resolution H\,$\alpha$ image from the Magellanic Clouds Emission Line Survey\,2 (MCELS2). 
The SNR shell (of angular diameter of $\approx2.7$ arcmin) is clearly visible in H\,$\alpha$ and 
[O\,{\sc iii}], but hardly can be detected in [S\,{\sc ii}]. This appears unusual because it is 
believed that in SNRs the intensity ratio of the [S\,{\sc ii}] $\lambda\lambda$6716, 6731 to H\,$\alpha$ 
emission lines should be quite large ($>0.4$; e.g. Mathewson \& Clarke 1973; Fesen, Blair \& Kirshner 1985). 
The low intensity of the [S\,{\sc ii}] emission lines was interpreted by H\'enault-Brunet et al. (2012) 
as the result of photoionization of the SNR shell and the local interstellar 
medium (ISM) by 2dFS\,3831 and/or by an ionizing radiation from hot massive stars in the star 
cluster NGC\,602 and a rich collection of OB stars within the giant shell SMC-SGS\,1 to the north of 
NGC\,602 (Fulmer et al. 2020). The SNR shell is most fully visible in the [O\,{\sc iii}] image, where its 
radius varies from $\approx75$ arcsec at the northwest rim to $\approx90$ arcsec in the opposite direction. 

MCSNR\,J0127$-$7332 was independently discovered by Haberl et al. (2012). In addition to the MCELS 
images of the SNR, they also presented its MOST (Molonglo Observatory Synthesis Telescope) 843\,MHz radio 
and {\it XMM-Newton} X-ray images. In radio the SNR shows a clear shell-like structure of the same 
size as the optical shell, while in X-rays it appears as a patchy diffuse emission confined within 
the optical shell. MCSNR\,J0127$-$7332 was also observed with the Australia Telescope Compact Array 
(ATCA) at $\approx1500$\,MHz (Haberl et al. 2012). By combining the ATCA and MOST data, Haberl et al. 
(2012) derived a spectral index of the SNR ($S_\nu \propto \nu ^\alpha$) of $\alpha=-0.8\pm0.4$, indicating 
the non-thermal nature of its radio emission.

Using the measured peak surface brightness of the northeast rim of MCSNR\,J0127$-$7332 and the 
apparent thickness of the SNR shell of is 5--10 per cent of the shell radius, H\'enault-Brunet et al. 
(2012) derived the number density of the shell and its mass of $1.3\pm0.3 \, {\rm cm}^{-3}$ and $250\pm100 
\, \msun$, respectively. Then, assuming that MCSNR\,J0127$-$7332 is in the Sedov phase, meaning that the 
kinetic energy of the shell is $\approx30$ per cent of the SN explosion energy (assumed to be equal to 
$E_0=10^{51}$\,erg), they derived the expansion velocity of the SNR and its age to be, respectively 
$V_{\rm SNR}=350\pm100 \, \kms$ and $t_{\rm SNR}=(2-4)\times10^4$\,yr. Similarly, assuming that the SNR 
is in the Sedov phase and using the temperature of X-ray emitting plasma of 0.23 keV (obtained from 
X-ray spectral modelling), Haberl et al. (2012) derived $t_{\rm SNR}\approx1.6\times10^4$\,yr, which 
implies $V_{\rm SNR}\approx440 \, \kms$.

\begin{table*}
\caption{Details of the SALT observations.}
\label{tab:log}
\begin{tabular}{llccccccc} \hline
Date & Grating & Exposure & Spectral scale & Spatial scale & PA & Slit & Seeing & Spectral range \\
  &  & (sec) & (\AA\,pixel$^{-1}$) & (arcsec\,pixel$^{-1}$) & ($\degr$) & (arcsec) & (arcsec) & (\AA) \\
  \hline
2012 October 13  & PG2300 & 400$\times$2  & 0.35  & 0.255 & 123  & 0.6  & 3.0 & 3817$-$4913  \\
2014 June 27     & PG2300 & 400$\times$2  & 0.34  & 0.255 & 123 & 1.5  & 1.3 & 3814$-$4904  \\
2014 July 09     & PG2300 & 400$\times$2  & 0.34  & 0.255 & 123  & 1.5  & 2.2 & 3814$-$4904  \\
2016 November 1  & PG2300 & 1100$\times$4 & 0.27  & 0.255 & 90  & 1.25 & 1.6 & 5900$-$6760  \\
2018 December 21 & PG2300 & 1500$\times$1 & 0.26  & 0.510 & 125  & 2.00 & 2.1 & 6070$-$6900  \\
2020 December 24 & PG900  & 1500$\times$1 & 0.97  & 0.255 & 90  & 1.25 & 1.2 & 3619$-$6708  \\
\hline
\end{tabular}
\end{table*}

\begin{figure}
\begin{center}
\includegraphics[width=8.5cm,angle=0]{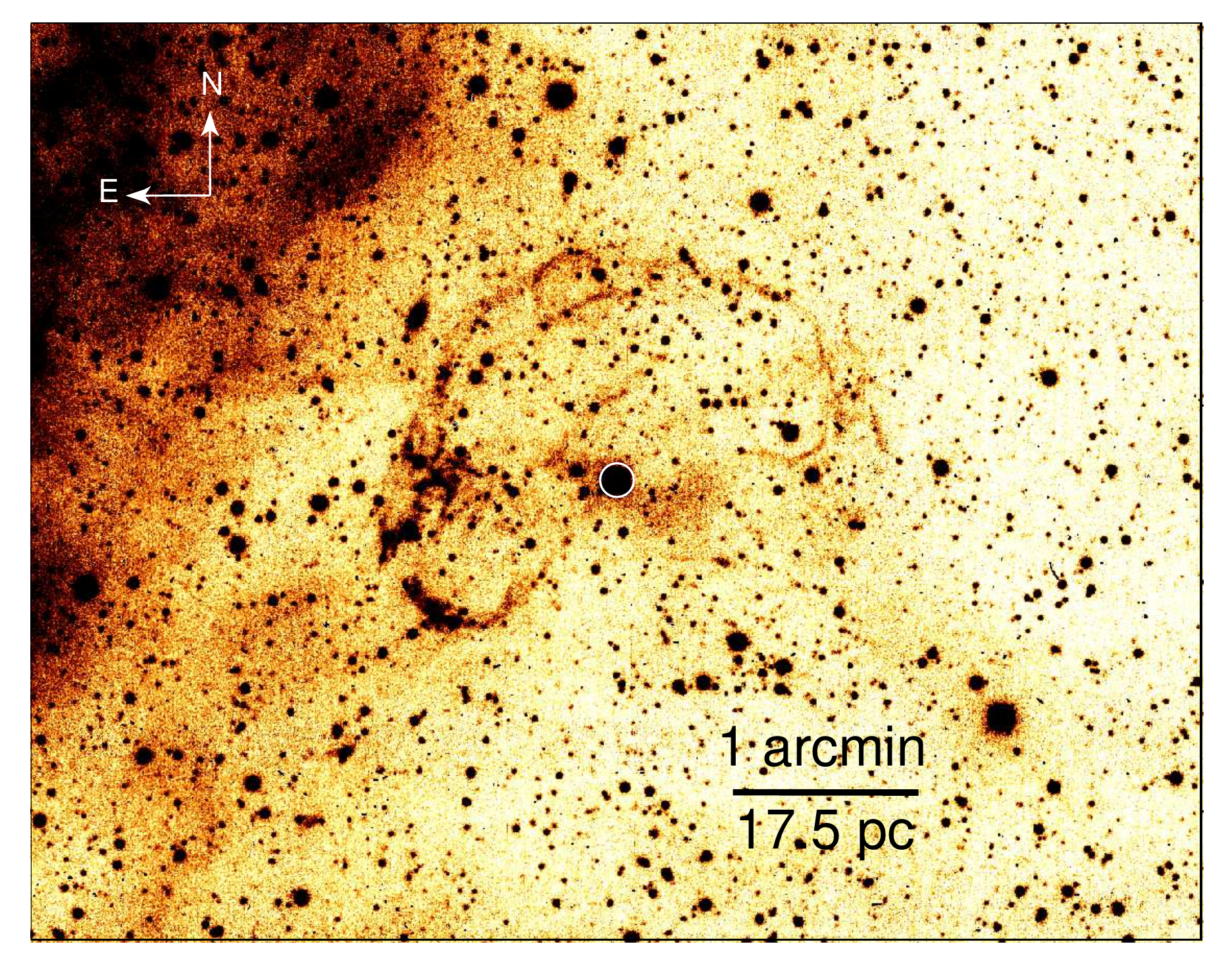}
\end{center}
\caption{MCELS2 H\,$\alpha$ image of MCSNR\,J0127$-$7332 and the mass donor star 2dFS\,3831 (marked 
with a white circle) of the BeXB SXP\,1062. The bright emission to the northeast of the SNR is the 
\hii region LHA\,115-N\,90 surrounding the massive star cluster NGC\,602.} 
\label{fig:smc}
\end{figure}

In Fig.\,\ref{fig:smc}, we show the MCELS2 H\,$\alpha$ image of MCSNR\,J0127$-$7332. The SNR 
appears as an incomplete almost circular filamentary shell. The bright star, 2dFS\,3831 (marked with a 
white circle), near the centre of the SNR is the mass donor of SXP\,1062. The east (bright) side of the SNR 
is faced towards the \hii region LHA\,115-N\,90 excited by the massive star cluster NGC\,602. This 
brightness asymmetry could be caused by increase in the number density of the local interstellar medium 
(ISM) towards the \hii region or by interaction of the SN blast wave with a gas outflow driven by massive 
stars in NGC\,602. In both cases, the blast wave would be somewhat slower in the east direction, which 
can explain why 2dFS\,3831 is offset from the geometric centre of the SNR towards its more bright edge.

In what follows, we assume that the SMC is located at 60 kpc (Hilditch, Howarth \& Harries 
2005). At this distance 1 arcmin corresponds to $\approx17.5$ pc. Correspondingly, the linear radius 
of the SNR is $R_{\rm SNR}\approx23$\,pc. 

\section{Observations}
\label{sec:obs}

We obtained long-slit spectra of MCSNR\,J0127$-$7332 with the Robert Stobie Spectrograph (RSS; Burgh 
et al. 2003; Kobulnicky et al. 2003) attached to the SALT (Buckley, Swart \& Meiring 2006; O'Donoghue 
et al. 2006). The observations were carried out in 2016, 2018 and 2020. In 2016, the spectra 
were obtained using the PG2300 grating with the spectral resolution FWHM of FWHM=$1.6\pm0.2$~\AA. 
In 2018, we used the same grating with a wider slit width, resulting in the spectral resolution FWHM of 
$2.2\pm0.2$~\AA. Hereafter, we will refer to these spectra as high-resolution spectra. 
In 2020 we obtained one more spectrum using the PG900 grating, which allowed us to cover a much wider 
spectral range, but with a lower spectral resolution FWHM of $4.5\pm0.5$~\AA \, (hereafter, a 
low-resolution spectrum). 

\begin{figure*}
\begin{minipage}[h]{0.49\linewidth}
\center{\includegraphics[width=1\linewidth,clip=0]{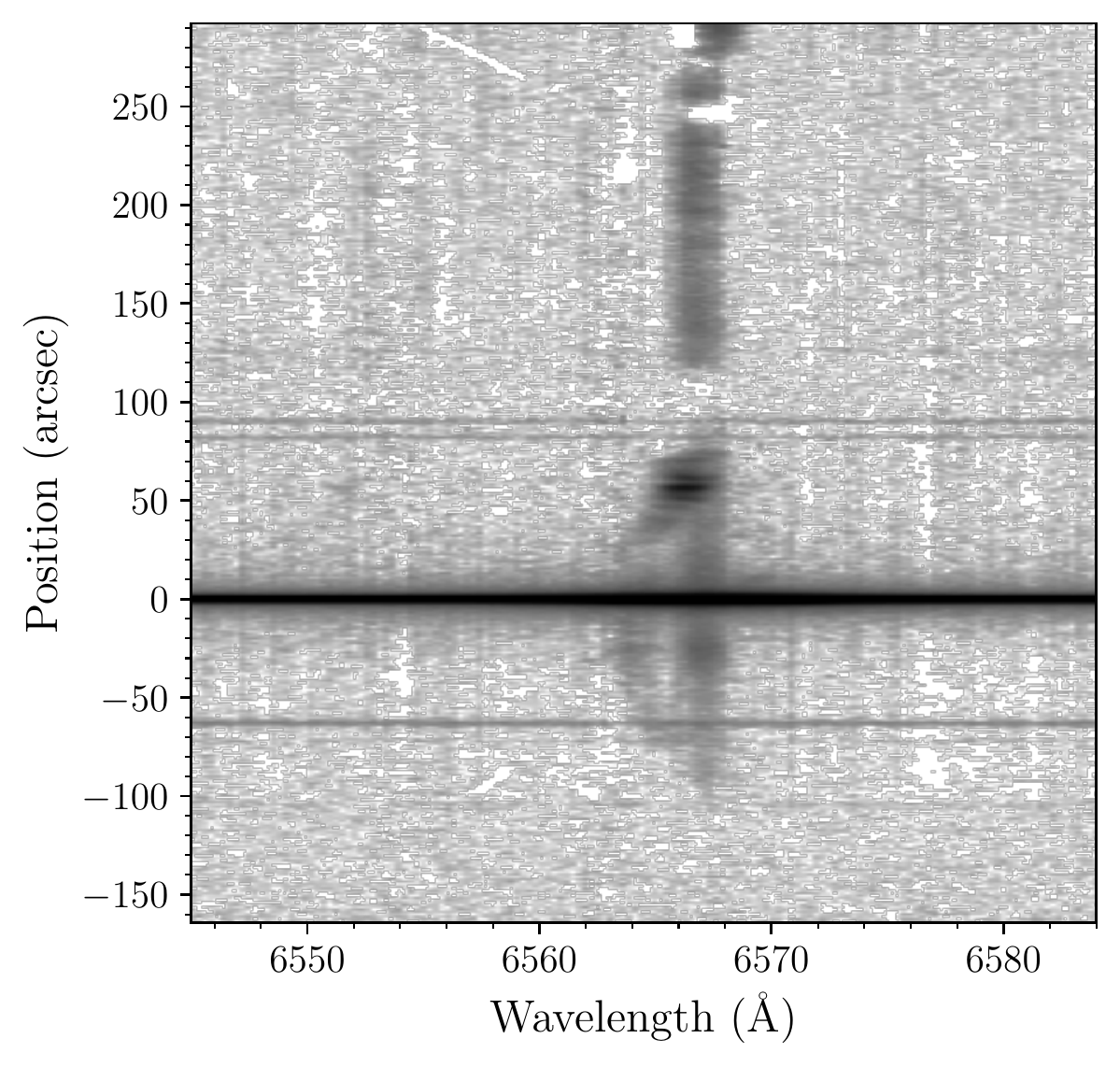}} {} \\
\end{minipage}
\hfill
\begin{minipage}[h]{0.49\linewidth}
\center{\includegraphics[width=1\linewidth,clip=0]{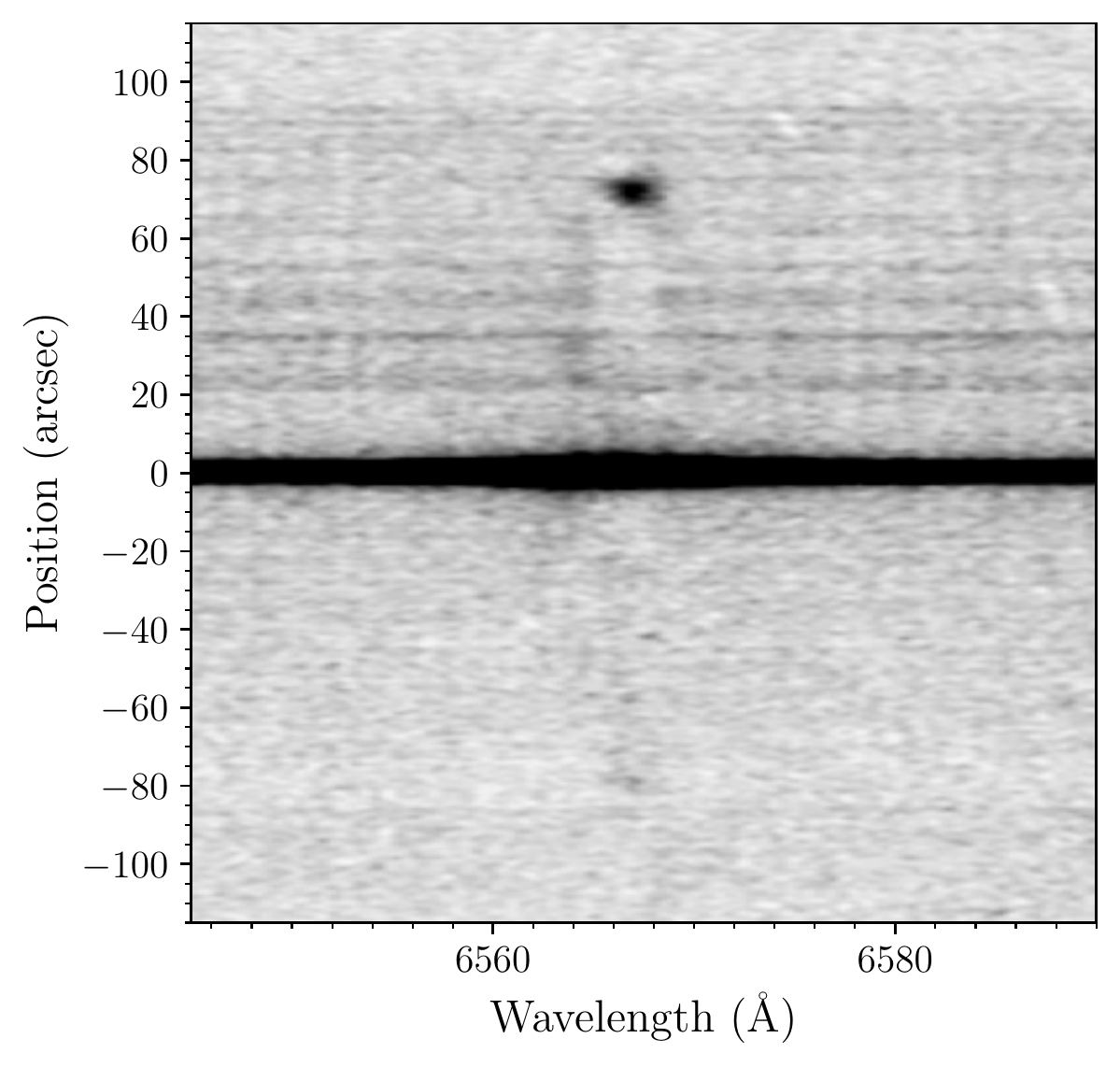}} {} \\
\end{minipage}
\vfill
\begin{minipage}[h]{0.49\linewidth}
\center{\includegraphics[width=0.9\linewidth,angle=270,clip=0]{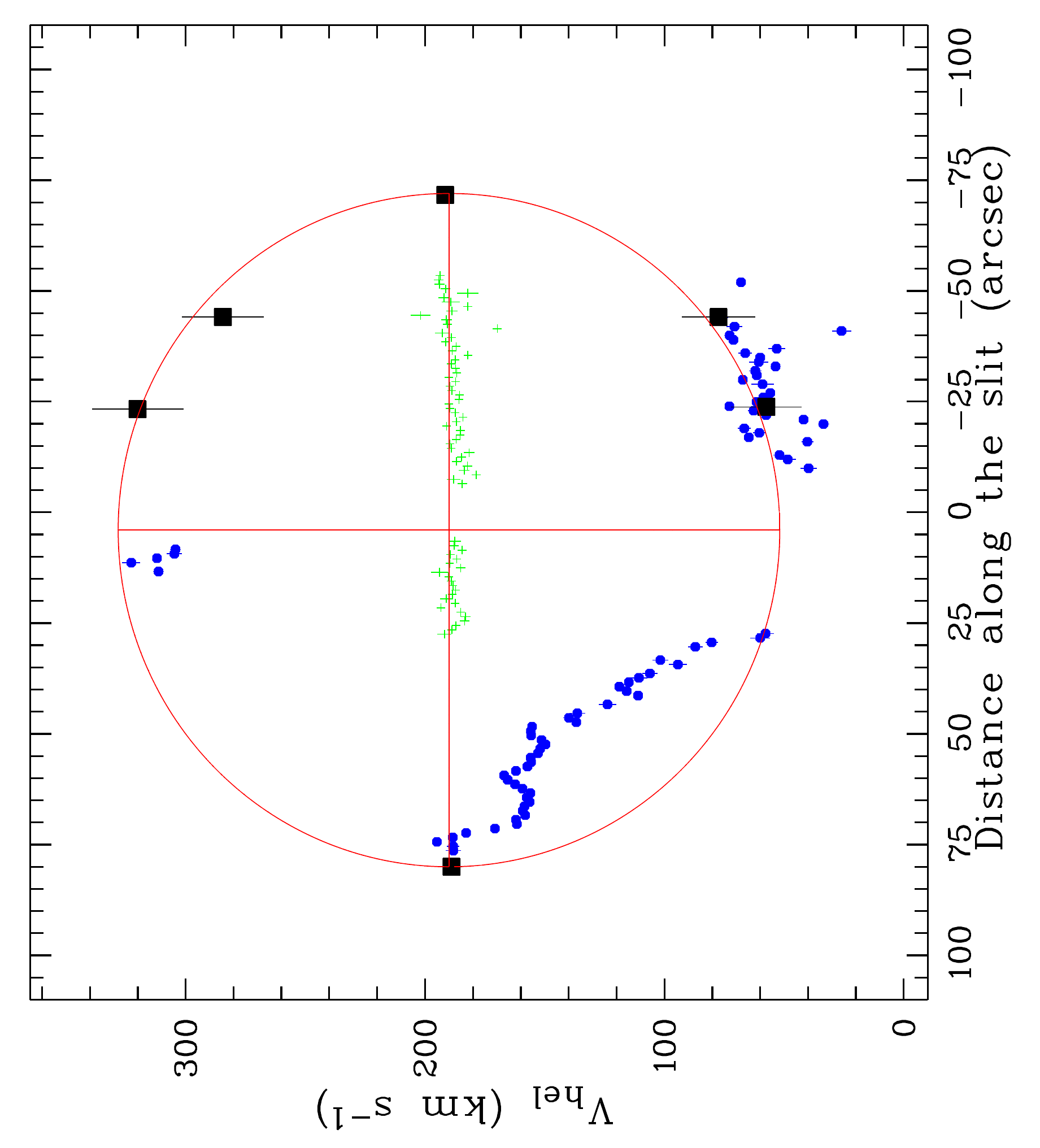}} {} \\
\end{minipage}
\hfill
\begin{minipage}[h]{0.49\linewidth}
\center{\includegraphics[width=1\linewidth,clip=0]{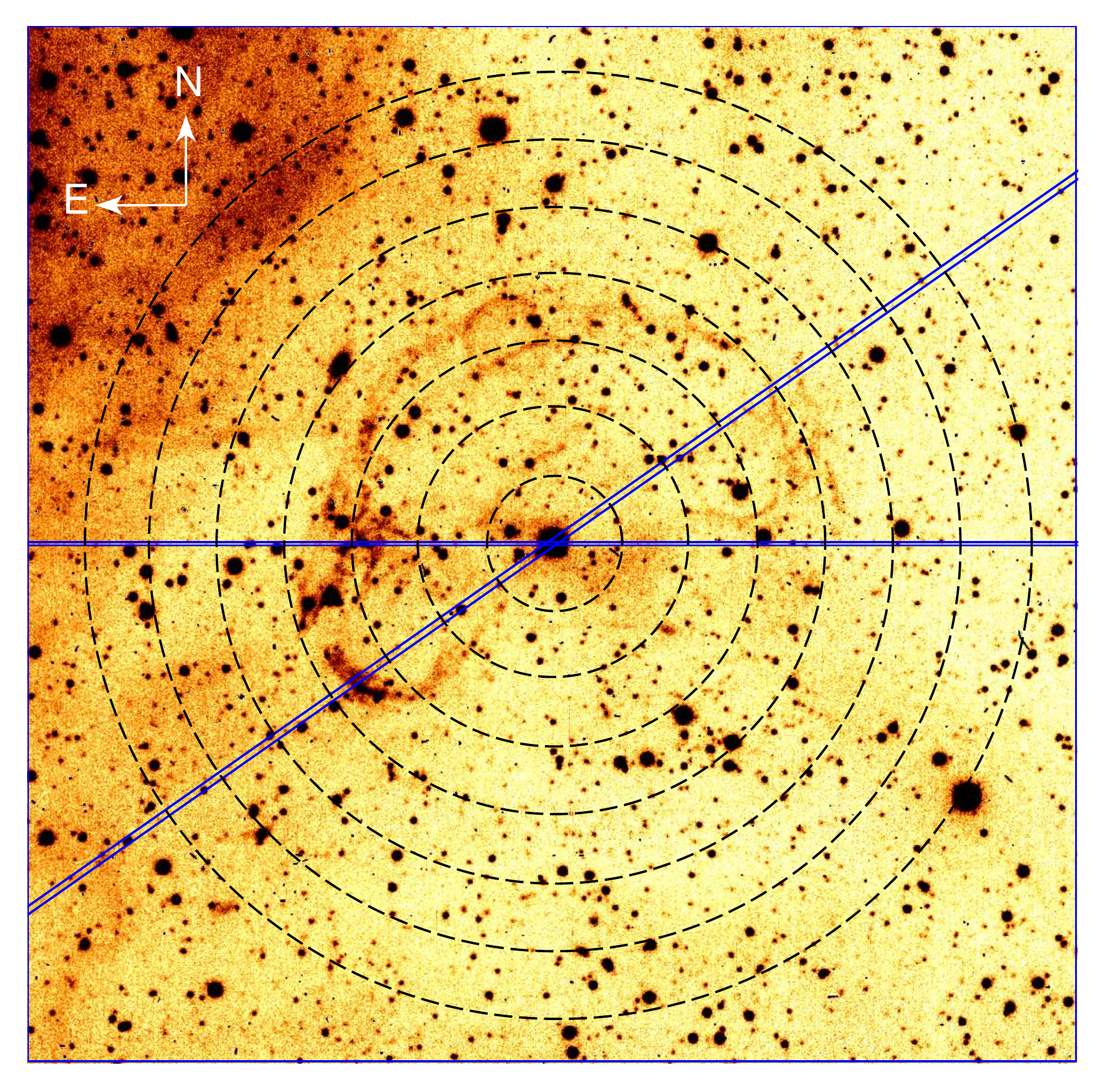}} {} \\
\end{minipage}
\caption{Upper panels: Portions of the 2D high-resolution spectra of the SNR shell, showing the 
H\,$\alpha$ emission line intensity along the slits with PA=90\degr \, (left-hand panel; east is up and 
west is down) and PA=125\degr \, (right-hand panel; southeast is up and northwest is down). Bottom 
left-hand panel: H\,$\alpha$ heliocentric radial velocity distribution along the slits. Black squares with 
error bars correspond to measurements at several positions in the 2D spectrum shown in the upper right
panel. The (blue) dots with error bars show the heliocentric radial velocity of the shell measured in 
the 2D spectrum shown in the upper left panel. The (green) crosses show the heliocentric radial velocity 
of the background H\,$\alpha$ emission of $\approx188\pm6 \, \kms$. Bottom right-hand panel: MCELS2 
H\,$\alpha$ image of the SNR MCSNR\,J0127$-$7332 centred on 2dFS\,3831. The positions of the SALT RSS 
slits with PA=90\degr \, and 125\degr \, are shown, respectively, by 1.25 and 2 arcsec wide blue 
rectangles. Concentric, dashed circles of angular radius in the range from 20 to 140 arcsec
with a step of 20 arcsec are added to the image to facilitate its comparison with the 2D spectra. See 
the text for details.
} 
\label{fig:vel}
\end{figure*}

In all these observations the slit was placed on 2dFS\,3831 and oriented in such a way as to cross 
the brightest knots in the eastern and southeastern edges of the shell (see Fig.\,\ref{fig:smc} and 
the bottom right panel of Fig.\,\ref{fig:vel}). Namely, in 2016 and 2020 the slit was oriented in the 
west-east direction, i.e. at a position angle (PA) of PA=90\degr (measured from north to east), while 
in 2018 it was oriented at PA=125\degr. The goal of these observations was to try to determine the 
expansion velocity of the SNR shell (as we did this for MCSNR\,J0103$-$7201; Gvaramadze et al. 2019) 
and to check whether the H\,$\alpha$ line in the spectrum of 2dFS\,3831 continues to change its EW. 
 
For wavelength calibration of the spectra an Xe lamp arc spectrum was taken immediately after the 
science frames. Spectrophotometric standard stars were observed at the same spectral setups for the 
relative flux calibration.

The obtained spectra were first reduced using the SALT science pipeline (Crawford et al. 2010), and 
further reduced as described in Kniazev et al. (2008). Calibration of the absolute flux is not possible 
with SALT because the telescope's unfilled entrance pupil moves during observations. Still, a relative 
flux correction to recover the spectral shape can be done using the observed spectrophotometric standards. 
 
We also extracted from the SALT archive three observational sets of 2dFS\,3831 obtained in 2012 and 2014. 
The spectra obtained in 2012 were presented in Sturm et al. (2013), while those obtained in 2014 
and 2016 were used in Gonz\'ales-Gal\'an et al. (2018) to study changes in EW(H\,$\alpha$) in the spectrum 
of 2dFS\,3831. For this paper, we reduced only blue parts of these spectra and used them to analyse 
2dFS\,3831 (see Section\,\ref{sec:star}). Since the red parts of the spectra cover only the spectral 
region around the H\,$\alpha$ line, in our analysis we just used published values of EW(H\,$\alpha$) from 
the literature. The spectral resolution FWHMs of the 2012 and 2014 spectra are $1.2\pm0.2$ and $2.0\pm0.3$~\AA,
respectively. 

Note that Sturm et al. (2013) and Gonz\'ales-Gal\'an et al. (2018; see their table~A1) give incorrect 
positional angles for their observations. The correct ones are given in Table\,\ref{tab:log} along with 
other details of all six observations utilized in this paper.

\begin{figure*}
\begin{center}
\includegraphics[width=14cm,angle=0,clip=]{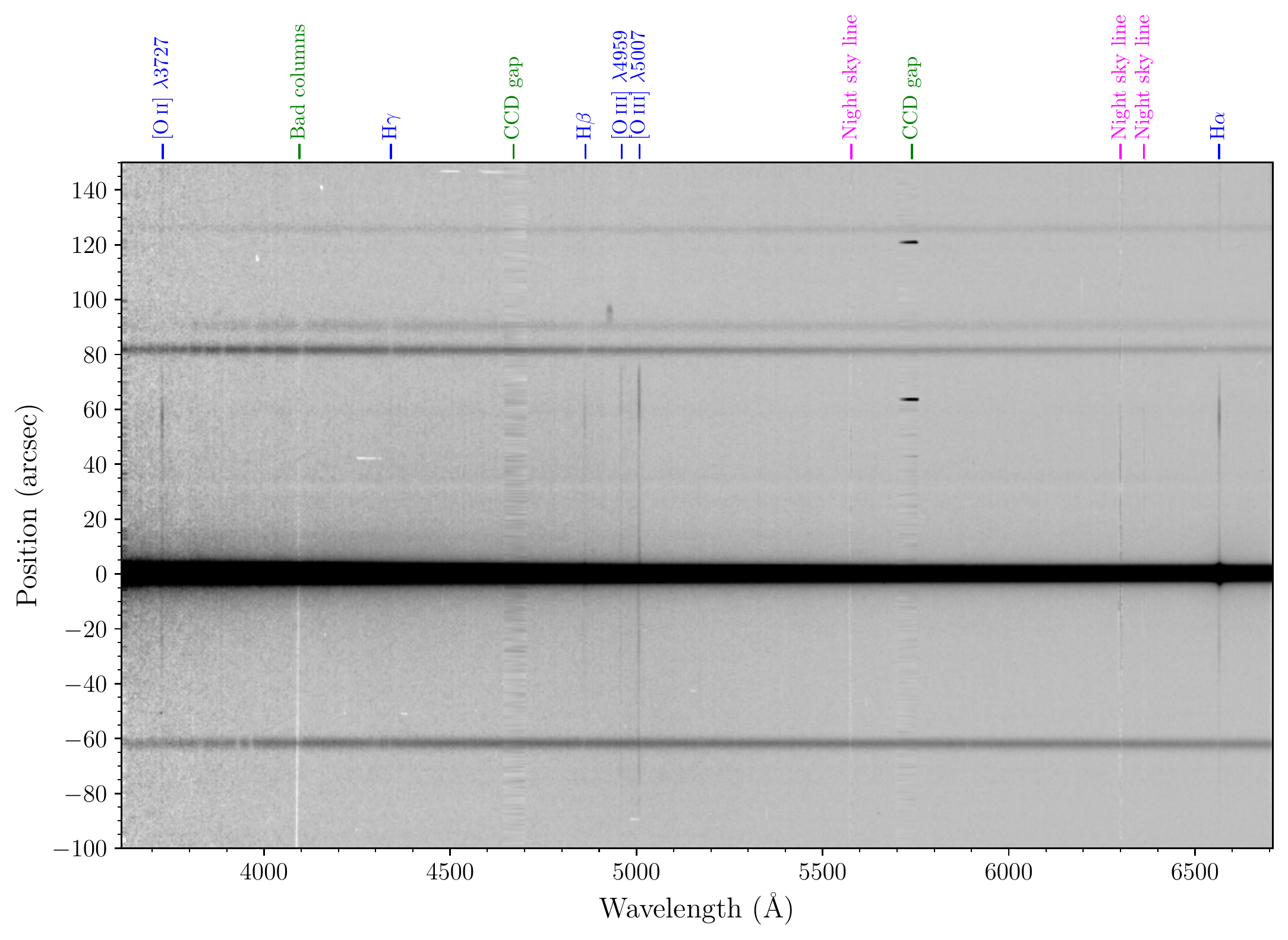}
\end{center}
\caption{2D low-resolution spectrum of MCSNR\,J0127$-$7332 and its environs. 
Positive offsets are east of 2dFS\,3831.}
\label{fig:2D}
\end{figure*}
%
\begin{figure}
\begin{center}
\includegraphics[width=8cm,angle=0,clip=]{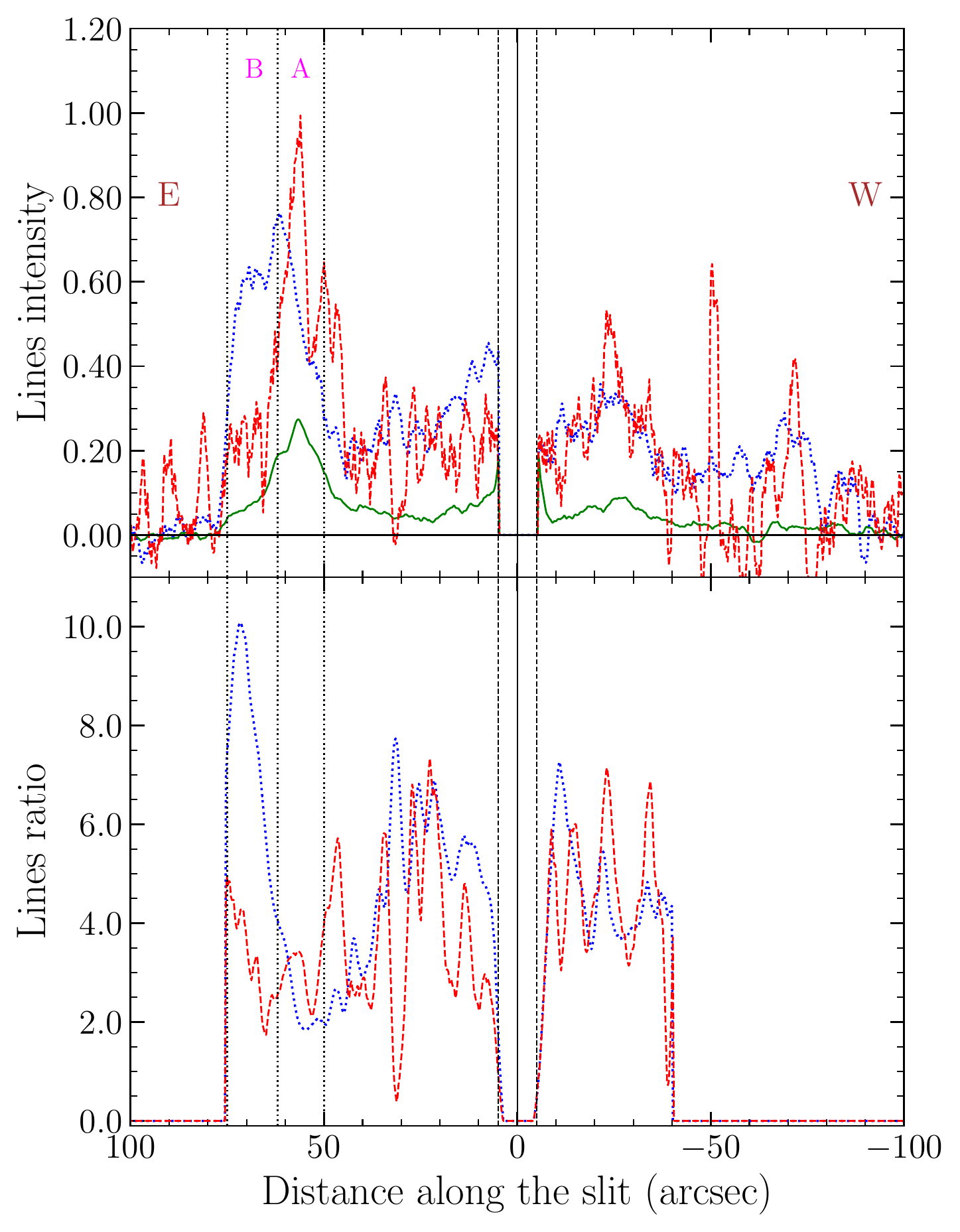}
\end{center}
\caption{Upper panel: [O\,{\sc iii}] $\lambda$5007, H\,$\beta$ and [O\,{\sc ii}] $\lambda$3727 
line intensity profiles along the slit for the low-resolution spectrum, shown, respectively, with blue 
(dotted), green (solid) and red (dushed) lines. The two vertical strips (bounded by dotted lines) show 
the regions (region\,A and region\,B) of the 2D spectrum from which the 1D spectra shown in 
Fig.\,\ref{fig:1D} were extracted.
Bottom panel: variations of the [O\,{\sc iii}]/H\,$\beta$ and [O\,{\sc ii}]/H\,$\beta$ line ratios along 
the slit, shown respectively by dotted (blue) and dashed (red) lines. The solid vertical line in both 
panels corresponds to the position of 2dFS\,3831, while the dashed vertical lines at $\pm5$ arcsec from 
the solid line mark the area where the nebular emission was not detected because of the effect of 
2dFS\,3831. E--W direction of the slit is shown.}
\label{fig:rat}
\end{figure}

\begin{figure*}
\begin{center}
\includegraphics[width=8.5cm,angle=0,clip=]{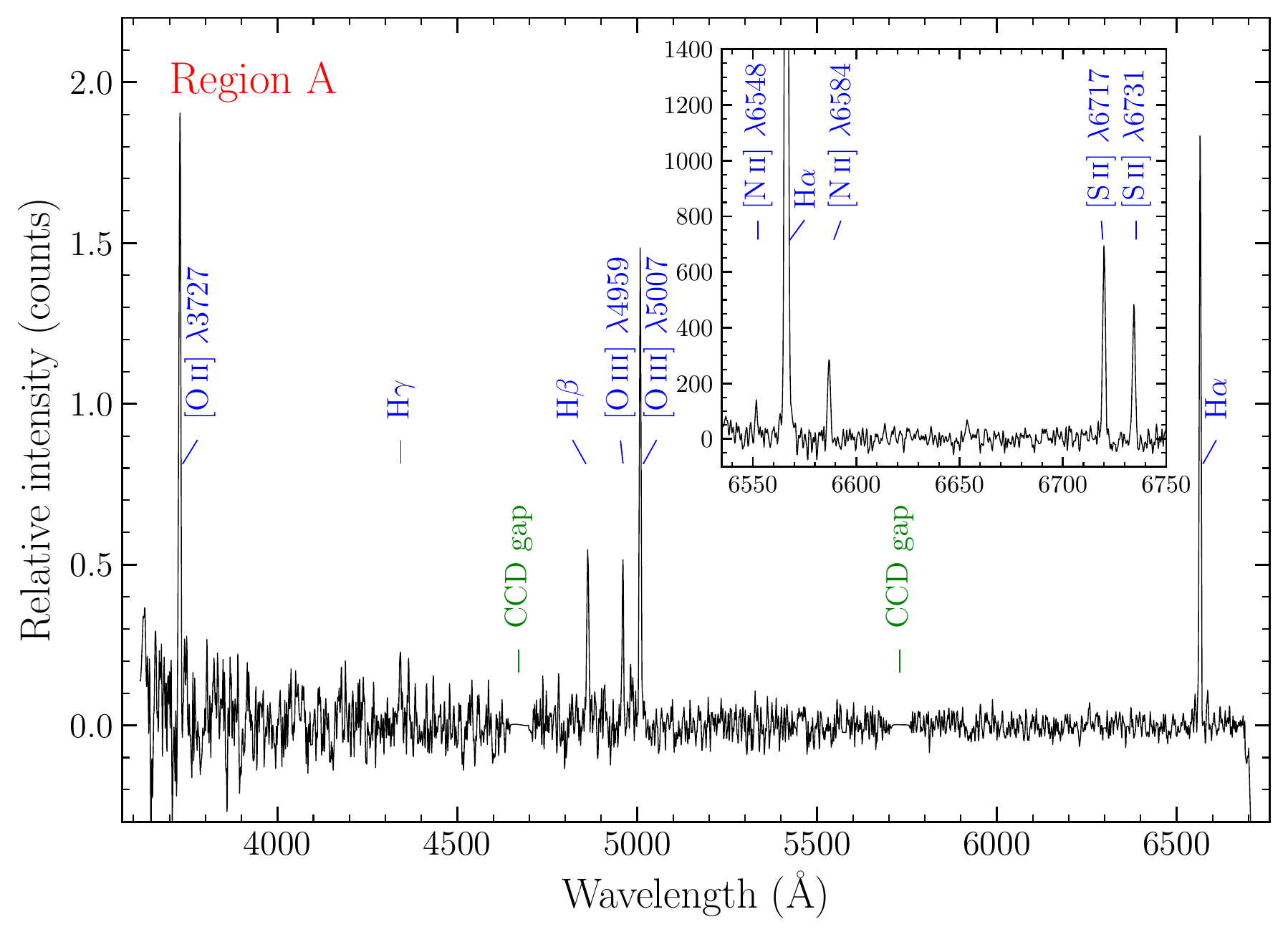}
\includegraphics[width=8.5cm,angle=0,clip=]{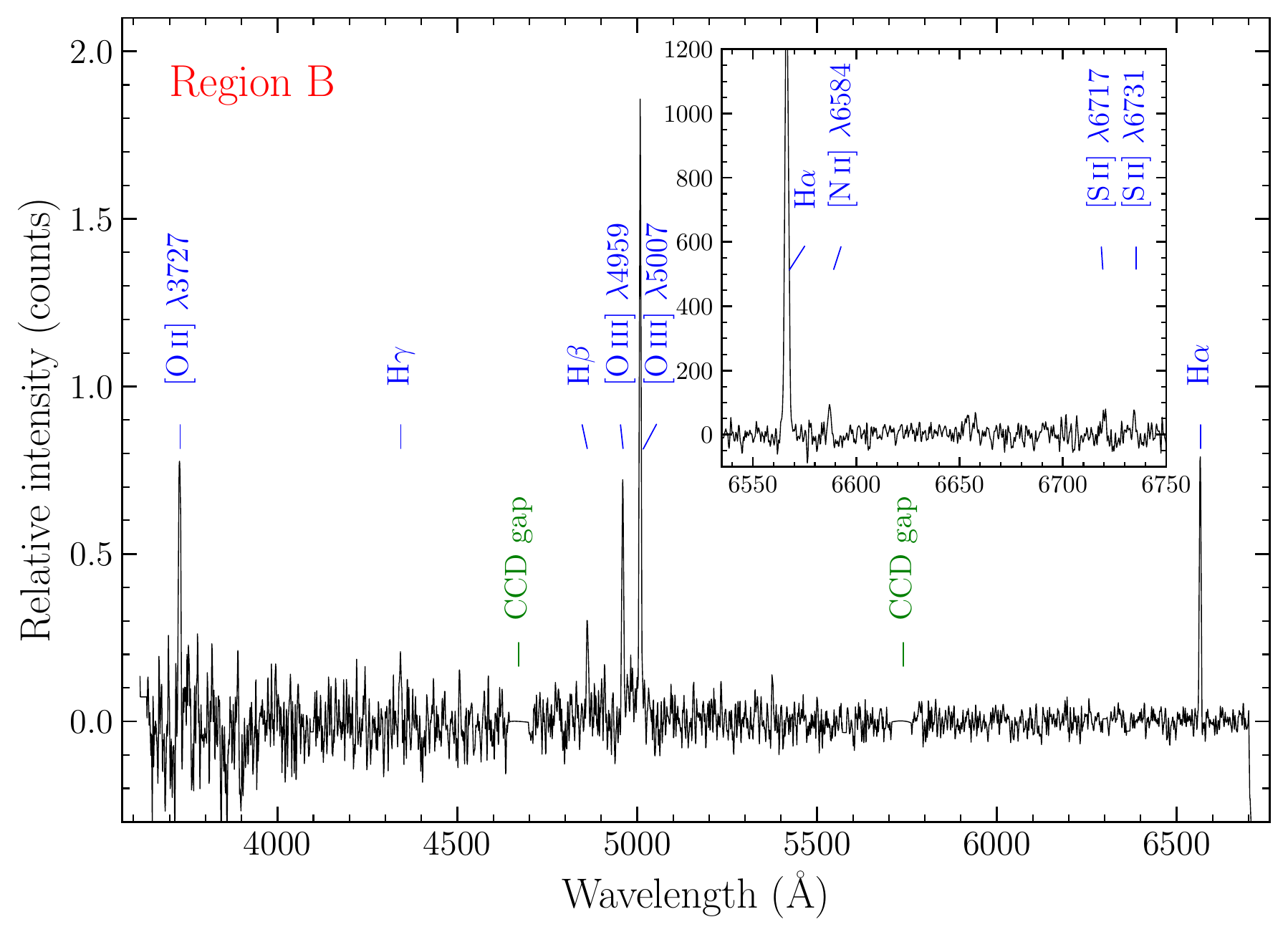}
\end{center}
\caption{1D low-resolution spectra of two regions in the east of the SNR. 
The inserts show portions of the high-resolution spectrum around the H\,$\alpha$ line. 
See text for details.}
\label{fig:1D}
\end{figure*}

\section{Results}
\label{sec:res}

\subsection{MCSNR\,J0127$-$7332}
\label{sec:shell}

In upper panels of Fig.\,\ref{fig:vel}, we present portions of the 2D high-resolution spectra of the SNR 
shell obtained for two slit orientations: PA=90\degr \, (left-hand panel) and PA=125\degr \, (right-hand 
panel). The left-hand panel shows that the H\,$\alpha$ emission along the slit comes from two major 
components: an almost straight (vertical) component (we attribute it to the background ISM; see below) and 
a blueshifted arc with a bright knot on its eastern edge (we attribute this component to the near side of 
the SNR shell). There is also a portion of redshifted (receding) side of the shell at $\approx+10$ arcsec.
The arc-like emission component extends along the slit between $\approx+75$ and $-90$ arcsec. The 
right-hand panel shows that at PA=125\degr \, we see both the receding and approaching sides of the shell. 
It also shows a bright knot at $\approx+75$ arcsec, which corresponds to the southeast edge of the SNR shell. 
In the northwest direction the shell extends to $\approx-80$ arcsec. Note that the second spectrum 
(PA=125\degr) was taken with an $\approx3$ times shorter exposure, an 1.6 time wider slit and about 
a factor of 1.3 worse seeing (see Table\,\ref{tab:log}). This explains why the background ISM emission 
is only weakly visible in this spectrum (e.g. at $+10$ arcsec).

To derive the expansion velocity of the SNR shell, $V_{\rm sh}$, we measured the heliocentric radial 
velocity, $V_{\rm hel}$, of the H\,$\alpha$ line along the slits using the 2D high-resolution spectra. The 
results of the measurements are plotted in the bottom left panel of Fig.\,\ref{fig:vel}. The black squares 
with error bars show heliocentric radial velocities measured at several positions in the 2D spectrum obtained 
with the slit oriented at PA=125\degr. One can see that these data points fit pretty well into the circle.
Assuming for the sake of simplicity that the SNR expands spherically symmetric, one finds 
$V_{\rm sh}\approx138\pm5 \, \kms$ and the systemic velocity of the shell of $V_{\rm sys}=190\pm5 \, \kms$. 

Similarly, we also measured $V_{\rm hel}$ using the second 2D high-resolution spectrum (PA=90\degr). These 
measurements are plotted in the same panel with (blue) dots with errors bars (in most cases the bars are 
shorter than the size of the dots). In general, they also fall well on the circle, except of data points 
in the range between $\approx+30$ and +70 arcsec, where $V_{\rm hel}$ shows systematically lower values. 
This deviation indicates that the east side of the shell expands with a somewhat lower velocity than the 
opposite one (cf. Section\,\ref{sec:obs}), which is consistent with the off-centred location of 2dFS\,3831 
within the SNR. It could also be caused in part by non-radial motions due to large-scale deformations of 
the shell, as evidenced by its complex structure on the east side from 2dFS\,3831 (see the bottom right 
panel of Fig.\,\ref{fig:vel}). The (green) crosses stretched horizontally from west to east correspond to 
the straight component of the H\,$\alpha$ emission in the 2D high-resolution spectrum with PA=90\degr. The 
mean $V_{\rm hel}$ of this component of $\approx188\pm6 \, \kms$ is equal to $V_{\rm sys}$ of the shell
and to $V_{\rm hel}$ of the \hii region to the east of the SNR (see the upper left and the bottom right 
panels of Fig.\,\ref{fig:vel}; cf. Nigra et al. 2008). Thus, we interpret this component as a background 
emission not related to the SNR. 

\begin{table}
\centering{
\caption{Observed line ratios in the low-resolution spectra of two regions in the east 
side of the SNR. See text for details.}
\label{tab:int}
\begin{tabular}{lcc} \hline
\rule{0pt}{10pt}
                                  & \MC{1}{c}{Region A}    & \MC{1}{c}{Region B}  \\ 
\rule{0pt}{10pt}
$\lambda_{0}$(\AA) Ion            & F($\lambda$)/F(H$\beta$)& F($\lambda$)/F(H$\beta$) \\ 
\hline
3727\ [O\ {\sc ii}]\              & 3.54$\pm$0.25  & 3.92$\pm$0.57  \\
4340\ H\,$\gamma$\                & 0.25$\pm$0.04  & 0.55$\pm$0.10  \\
4861\ H\,$\beta$\                 & 1.00$\pm$0.09  & 1.000$\pm$0.16 \\
4959\ [O\ {\sc iii}]\             & 0.67$\pm$0.06  & 2.90$\pm$0.35  \\
5007\ [O\ {\sc iii}]\             & 2.42$\pm$0.16  & 7.99$\pm$0.92  \\
6300\ [O\ {\sc i}]\               & 0.07$\pm$0.01  & ---            \\
6548\ [N\ {\sc ii}]\              & 0.06$\pm$0.01  & ---            \\
6563\ H\,$\alpha$\                & 2.98$\pm$0.19  & 3.19$\pm$0.36  \\
6584\ [N\ {\sc ii}]\              & 0.17$\pm$0.01  & 0.15$\pm$0.03  \\
6717\ [S\ {\sc ii}]\              & 0.42$\pm$0.03  & 0.12$\pm$0.03  \\
6731\ [S\ {\sc ii}]\              & 0.30$\pm$0.02  & 0.11$\pm$0.02  \\
& & \\
$\log$([O\ {\sc ii}]/H\,$\beta$)  & $ 0.55\pm0.05$ & $ 0.59\pm0.09$ \\
$\log$([O\ {\sc iii}]/H\,$\beta$) & $0.38\pm0.05$  & $0.90\pm0.08$  \\
$\log$([O\ {\sc  i}]/H\,$\alpha$) & $-1.61\pm0.06$ & ---            \\
$\log$([N\ {\sc ii}]/H\,$\alpha$) & $-1.11\pm0.04$ & $-1.21\pm0.09$ \\
$\log$([S\ {\sc ii}]/H\,$\alpha$) & $-0.62\pm0.04$ & $-1.15\pm0.09$ \\
\hline
\end{tabular}
}
\end{table}

Fig.\,\ref{fig:2D} shows a portion of the 2D low-resolution spectrum of MCSNR\,J0127$-$7332 
and its environs. In this spectrum the SNR occupies the area between $\approx-90$ and +75 arcsec
(positive offsets are east of 2dFS\,3831). One can see that the spectrum of the SNR is dominated by 
emission lines of H\,$\alpha$, H\,$\beta$, [O\,{\sc iii}] $\lambda\lambda$4959, 5007 and [O\,{\sc ii}] 
$\lambda$3727 (a blend of the [O\,{\sc ii}] $\lambda\lambda$3726, 3729 lines), whose intensities are 
maximum near the eastern edge of the SNR shell. One can also see that the high-excitation [O\,{\sc iii}] 
emission lines are confined within boundaries of the SNR, while the [O\,{\sc ii}] and H\,$\alpha$ lines 
are also visible beyond the SNR at angular distances greater than $+120$ arcsec, i.e. in the \hii region 
to the east of the SNR (compare with the upper left and bottom right panels of Fig.\,\ref{fig:vel}).  

The upper panel of Fig.\,\ref{fig:rat} plots the [O\,{\sc iii}] $\lambda$5007, H\,$\beta$ and [O\,{\sc ii}] 
line intensity profiles along the slit. It shows that the intensities of the H\,$\beta$ and [O\,{\sc ii}] 
lines are maximum at the position of the bright knot near the eastern edge of the SNR shell (the region 
between +50 and +62 arcsec from 2dFS\,3831; see also the bottom right panel in Fig.\,\ref{fig:vel}) and 
drop to zero at the edge of the SNR. In contrast to these lines, the intensity of the [O\,{\sc iii}] 
$\lambda$5007 line reaches its maximum at the outer border of the knot and remains high to the very edge 
of the shell. Correspondingly, the [O\,{\sc iii}] to H\,$\beta$ line ratio abruptly increases by a factor 
of about 5 beyond the knot, indicating high excitation conditions in this part of the SNR (see the bottom 
panel of Fig.\,\ref{fig:rat}).

Fig.\,\ref{fig:1D} presents 1D spectra of two regions in the east of the SNR. These spectra were extracted 
from the 2D low-resolution spectrum by summing up rows in the ranges from +50 to +62 arcsec (hereafter 
region\,A) and from +62 to +75 arcsec (hereafter region\,B). The region\,A coincides with the bright knot 
(see Fig.\,\ref{fig:vel}), while the region\,B corresponds to the high-excitation region to the east 
of the knot. The inserts in Fig.\,\ref{fig:1D} show portions of the 1D high-resolution spectrum (PA=90$\degr$) 
around the H\,$\alpha$ line. Besides the H\,$\alpha$, H\,$\beta$, [O\,{\sc iii}] and [O\,{\sc ii}] emission 
lines, the spectra also show much weaker emission lines due to H\,$\gamma$, [N\,{\sc ii}] $\lambda$6584, and 
[S\,{\sc ii}] $\lambda\lambda$6716, 6731. In the high-resolution spectrum of region\,A, we also detected the 
[O\,{\sc i}] $\lambda$6300 (not shown in Fig.\,\ref{fig:1D}) and [N\,{\sc ii}] $\lambda$6548 emission lines. 
All detected lines were measured using programs described in Kniazev et al. (2004) and their observed 
intensities normalized to H\,$\beta$, $F(\lambda)/F$(H\,$\beta$), are given in Table\,\ref{tab:int}. This 
table also gives several diagnostic emission line ratios, which can be used to separate SNRs from \hii regions 
and circumstellar nebulae. In these ratios, the symbols [N\ {\sc ii}] and [S\ {\sc ii}] mean the sum of the 
intensities of both lines of the doublet. 

From Table\,\ref{tab:int} it follows that the intensity ratio of the combined [S\,{\sc ii}] lines against 
H\,$\alpha$ of $\approx0.07-0.24$ is below the low end of a range of values ($\geq0.3-0.4$) commonly used 
to separate SNRs from other types of emission nebulae (e.g., Rosado et al. 1983; Georgelin et al. 1983; Frew 
\& Parker 2010; Leonidaki, Boumis \& Zezas 2013). On the other hand, Kopsacheili, Zezas \& Leonidaki (2020) 
showed that the use of the [S\,{\sc ii}]/H\,$\alpha >0.4$ criterion leads to a selection effect against 
evolved SNRs, because the [S\,{\sc ii}]/H\,$\alpha$ ratio produced by their low-velocity ($\sim100 \, \kms$) 
shocks is predominantly $<0.4$ (see also Dopita \& Sutherland 1996). Moreover, Kopsacheili et al. (2020) found 
that at subsolar metallicities the use of the [S\,{\sc ii}]/H\,$\alpha >0.4$ criterion may lead to rejection 
of about 70 per cent of genuine SNRs. Instead, they proposed to use 2D and 3D diagnostics based on intensity 
ratios between optical forbidden lines (the ones that are usually stronger in shock-excited nebulae than in 
photoionized ones) and their closest Balmer lines. Namely, in addition to the [S\,{\sc ii}] 
$\lambda\lambda$6717, 6731 lines, it was proposed to use the [N\,{\sc ii}] $\lambda$6584, [O\,{\sc i}] 
$\lambda$6300, [O\,{\sc ii}] $\lambda$3727 and [O\,{\sc iii}] $\lambda$5007 lines as well. 

To create diagnostic diagrams, Kopsacheili et al. (2020) used theoretical line ratios from grids of 
shock-excitation and photoionization models (MAPPINGS III) by Allen et al. (2008), which were calculated 
for wide ranges of shock velocities (from 100 to $1000 \, \kms$) and magnetic parameters $B_{\rm ISM}/n_{\rm 
ISM} ^{1/2}$ (from $10^{-4}$ to $10 \,\mu{\rm G} \, {\rm cm}^{3/2}$), where $B_{\rm ISM}$ and $n_{\rm ISM}$ 
are, respectively, the transverse component of the local ISM magnetic field and the local ISM number density, 
and different abundances (including that of the SMC). Particularly, their 
[S\,{\sc ii}]/H\,$\alpha$--[O\,{\sc i}]/H\,$\beta$ diagnostic diagram (see their fig.\,5) shows that there 
is a significant fraction of shock models with [S\,{\sc ii}]/H\,$\alpha <0.4$. Using line ratios from 
Table\,\ref{tab:int}, one can see that MCSNR\,J0127$-$7332 falls in the SNR locus in all diagnostic diagrams 
constructed by Kopsacheili et al. (2020; see their figs 4, 5 and 7--11), which supports the classification of 
MCSNR\,J0127$-$7332 as a SNR.

\subsection{2dFS\,3831}
\label{sec:star}

The obtained spectra of 2dFS\,3831 were analysed with the {\sc fbs} (Fitting Binary Stars) software 
(Kniazev et al. 2020; Katkov et al., in preparation). This software allows to determine parameters of 
individual components of binary systems such as effective temperature $T_{\rm eff}$, surface gravity 
$\log g$, projected rotational velocity $v\sin i$ (where $v$ is the equatorial rotational velocity and 
$i$ is the inclination angle between the rotational axis and the line-of-sight), metallicity [Fe/H], 
and heliocentric radial velocity $V_{\rm hel}$, as well as the colour excess $E(B-V)$ of the system. 
{\sc fbs} simultaneously approximates the observed spectrum by a model, which is obtained by 
interpolating over the grid of theoretically calculated high-resolution stellar spectra, and convolves 
it with a function that takes into account the broadening and wavelength shift of lines caused,
respectively, by stellar rotation and motion along the line-of-sight at a given epoch. In case of a 
single star or a binary with a degenerate companion, the fitting routine uses one model spectrum for 
the single/non-degenerate star. 

\begin{figure*}
\begin{center}
\includegraphics[width=8.5cm,angle=0,clip=]{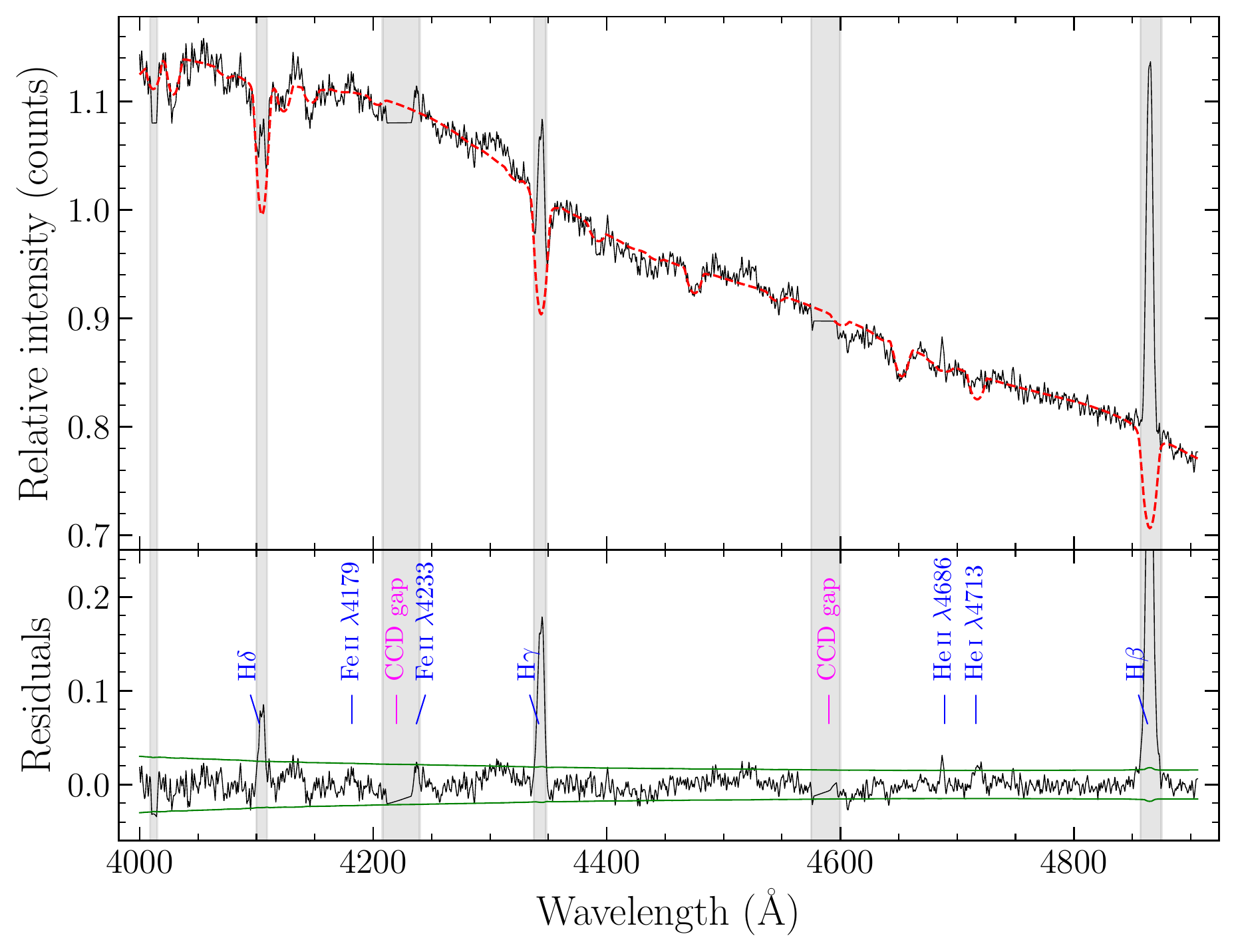}
\includegraphics[width=8.5cm,angle=0,clip=]{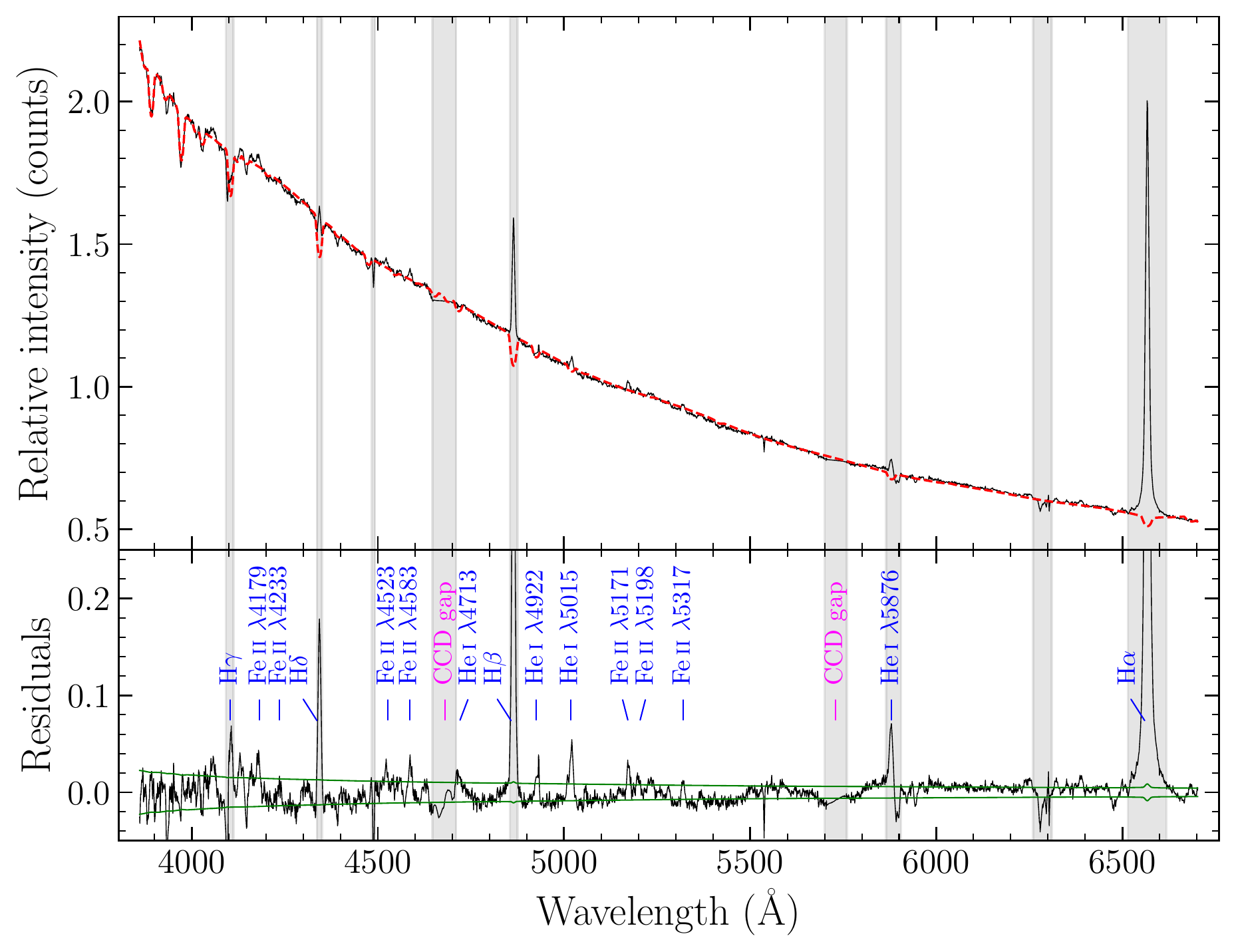}
\end{center}
\caption{Results of modelling of two spectra of 2dFS\,3831 obtained on 2014 June 17 (left-hand 
panel) and 2020 December 24 (right-hand panel). Upper panels: comparison of the observed spectra 
(solid black lines) with the best-fitting models (red dashed lines) obtained with the {\sc fbs} 
software. The grey vertical areas show spectral regions excluded from the spectral fit because of 
their contamination by emission from the circumstellar disk. Bottom panels: the difference between 
the observed and model spectra (black noisy line). The green solid lines indicate $1\sigma$ errors in the 
observed spectrum. Note that weak emission lines became more noticeable after the model subtraction. 
The positions of identified emission lines and CCD gaps are indicated.} 
\label{fig:star}
\end{figure*}

We separately fitted the available four wide-range spectra of 2dFS\,3831 with the synthetic spectra from 
the {\sc tlusty} models (Lanz \& Hubeny 2003, 2007) that were convolved to the spectral resolution of each 
particular observation. The results of the fit of two of them are presented in Fig.\,\ref{fig:star}, while 
mean values of the parameters determined from the fit of all four spectra are listed in Tables\,\ref{tab:par}.  
Note that the obtained value of metallicity of 2dFS\,3831 agrees fairly well with the metallicity
of the SMC of $-1.25\pm0.01$\,dex (Cioni 2009). 
Note also that the rotational velocity estimate should be considered with caution because it was obtained 
from the low-resolution spectra and because {\sc fbs} does not take into account the line broadening 
due to macroturbulence. On the other hand, the independent estimate of $v\sin i$ of $450 \, \kms$ based on
better spectroscopic data and the state-of-the-art stellar atmosphere models (Ramachandran et al. 2019) 
is not much less than our estimate. Since accounting for the effect of macroturbulence cannot significantly 
reduce these velocity estimates (e.g. Grassitelly et al. 2016), it is reasonable to assume that the axis of 
rotation of the circumstellar disk is tilted at a significant angle to our line-of-sight, which is at variance 
with the suggestion by Gonz\'alez-Gal\'an et al. (2018) that the disk is oriented face-on.

\begin{table}
\centering\caption{Parameters of 2dFS\,3831.}
\begin{tabular}{lc}
\hline
Parameter  & Value \\
\hline
$T_{\rm eff}$ (K)       & $25\,000\pm770$ \\                     
$\log g$                & $2.7\pm0.2$ \\                      
$[$Fe/H$]$              & $-1.40\pm0.04$ \\                      
$v\sin i \, (\kms)$     & $580\pm60$ \\                      
$E(B-V)$ (mag)          & $0.27\pm0.05$ \\                       	                   
\hline
\end{tabular}
\label{tab:par}
\end{table}

\begin{table*}
\centering\caption{Changes in EW(H\,$\alpha$) with the orbital phase in the spectra 
of 2dFS\,3831.}
\label{tab:EW}
\renewcommand{\footnoterule}{}
\begin{center}
\begin{tabular}{lcrcc} 
\hline
Date             & MJD & EW(\AA) & Reference & Phase \\
\hline
1999 September 15& 51436 & $-23\pm1$     & H\'enault-Brunet et al. (2012) & 0.194 \\
2012 October 13  & 56214 & $-26.7\pm0.1$ & Sturm et al. (2013)            & 0.089 \\
2014 June 20     & 56829 & $-36.3\pm0.6$ & Gonz\'alez-Gal\'an et al. 2018 & 0.026 \\
2014 June 27     & 56835 & $-34.0\pm0.6$ & Gonz\'alez-Gal\'an et al. 2018 & 0.036 \\
2014 July 9      & 56847 & $-34.1\pm0.6$ & Gonz\'alez-Gal\'an et al. 2018 & 0.054 \\
2016 November 2  & 57694 & $-33\pm1$     & Gonz\'alez-Gal\'an et al. 2018 & 0.345 \\
2018 December 21 & 58474 & $-31.8\pm0.5$ & this work                      & 0.535 \\
2020 December 24 & 59207 & $-34.7\pm0.2$ & this work                      & 0.654 \\ 
\hline
\end{tabular}
\end{center}
\end{table*}

Using the last two spectra of 2dFS\,3831 we measured EW(H\,$\alpha$) and found that it was equal 
to $-31.8\pm0.5$\,\AA \, and $-34.7\pm0.2$\,\AA \, in 2018 and 2020, respectively. A comparison of 
these values with the EWs measured in 1999--2016 (see Table\,\ref{tab:EW}) shows that the absolute 
value of EW(H\,$\alpha$) has reached a low in 2018 and then began to increase again. The changes in 
EW may reflect changes in the size/geometry of the circumstellar (excretion) disk around 2dFS\,3831 
caused by the feedback from the companion neutron star (e.g. Reig, Fabregat \& Coe 1997) and/or by 
the variable mass loss from the Be star (e.g. Rajoelimanana, Charles \& Udalski 2011). 

\begin{figure*}
\begin{center}
\includegraphics[width=8.5cm,angle=0]{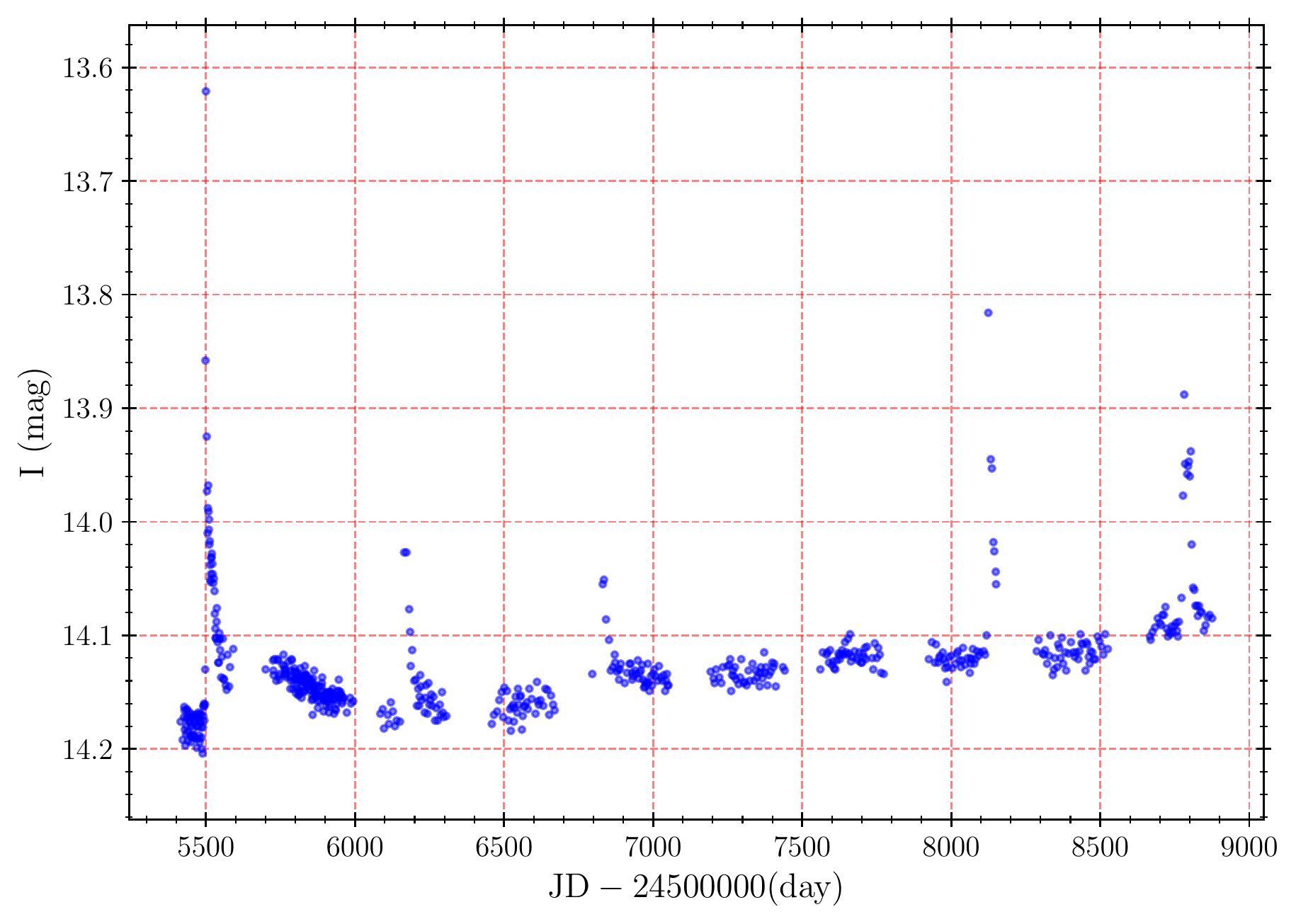}
\includegraphics[width=8.5cm,angle=0]{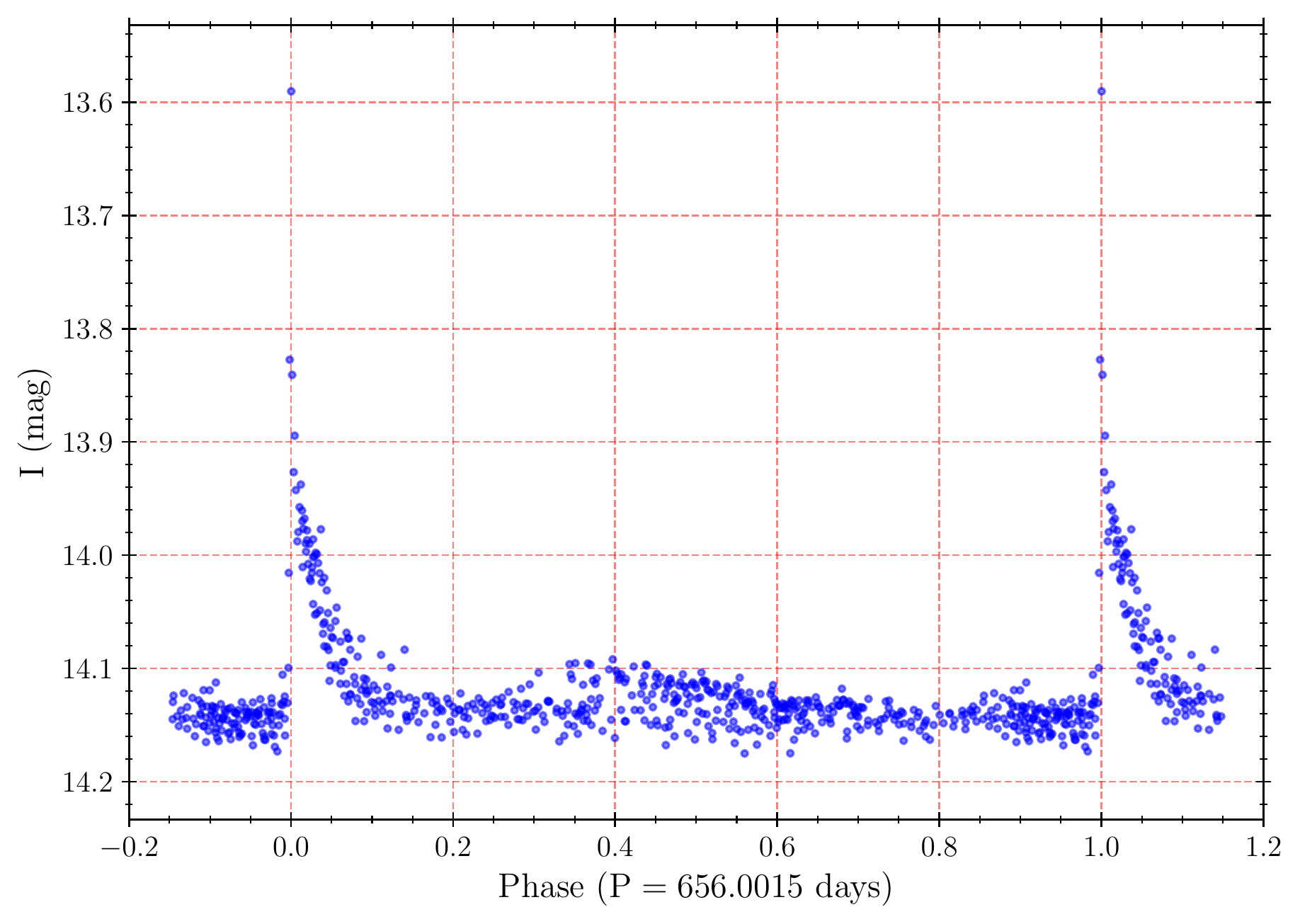}
\end{center}
\caption{Left-hand panel: OGLE $I$-band light curve of 2dFS\,3831 in 2010--2020. Note the 
gradual increase of the quiescent $I$-magnitude with time. Right-hand panel: The light curve folded 
with a period of 656.0015\,d and corrected for the linear increase of the quiescent $I$-magnitude 
(see text for details).}
\label{fig:LC}
\end{figure*}

To search for possible correlation between the changes 
in EW(H\,$\alpha$) and the binary orbital period $P_{\rm orb}$, we first re-evaluated $P_{\rm orb}$
using the latest light curve of 2dFS\,3831 from the Optical Gravitational Lensing 
Experiment\footnote{http://ogle.astrouw.edu.pl/ogle4/xrom/xrom.html} (OGLE; Udalski 2008). This light 
curve covers the time period from 2010 August 6 to 2020 January 27 (see the left-hand panel of 
Fig.\,\ref{fig:LC}), during which the system has experienced six outbursts, one of which was not covered 
because it fell in the gap between observations. Note that the light curve shows a clear trend of gradual 
increase of the quiescent $I$-magnitude with time (the possible existence of this trend was previously 
pointed out by Gonz\'alez-Gal\'an et al. 2018). After correction of the OGLE photometry for this trend 
(interpolated with a first-order polynomial function), we used the method from Lafler \& Kinman (1965), 
that was implemented for our project of study of long-period eclipsing binaries (Kniazev et al. 2020), 
to derived $P_{\rm orb}=656.0015\pm0.0010$\,d with the epoch of maximum light at JD 
$2455499.5823\pm0.0015$. The obtained result is in excellent agreement with the orbital ephemerides of 
SXP\,1062 from Schmidtke et al. (2019), which are based on the OGLE light curve covering the first four 
observed outbursts. The light curve folded with $P_{\rm orb}$ is shown in the right-hand panel of 
Fig.\,\ref{fig:LC}.

\begin{figure}
\begin{center}
\includegraphics[width=8.5cm,angle=0]{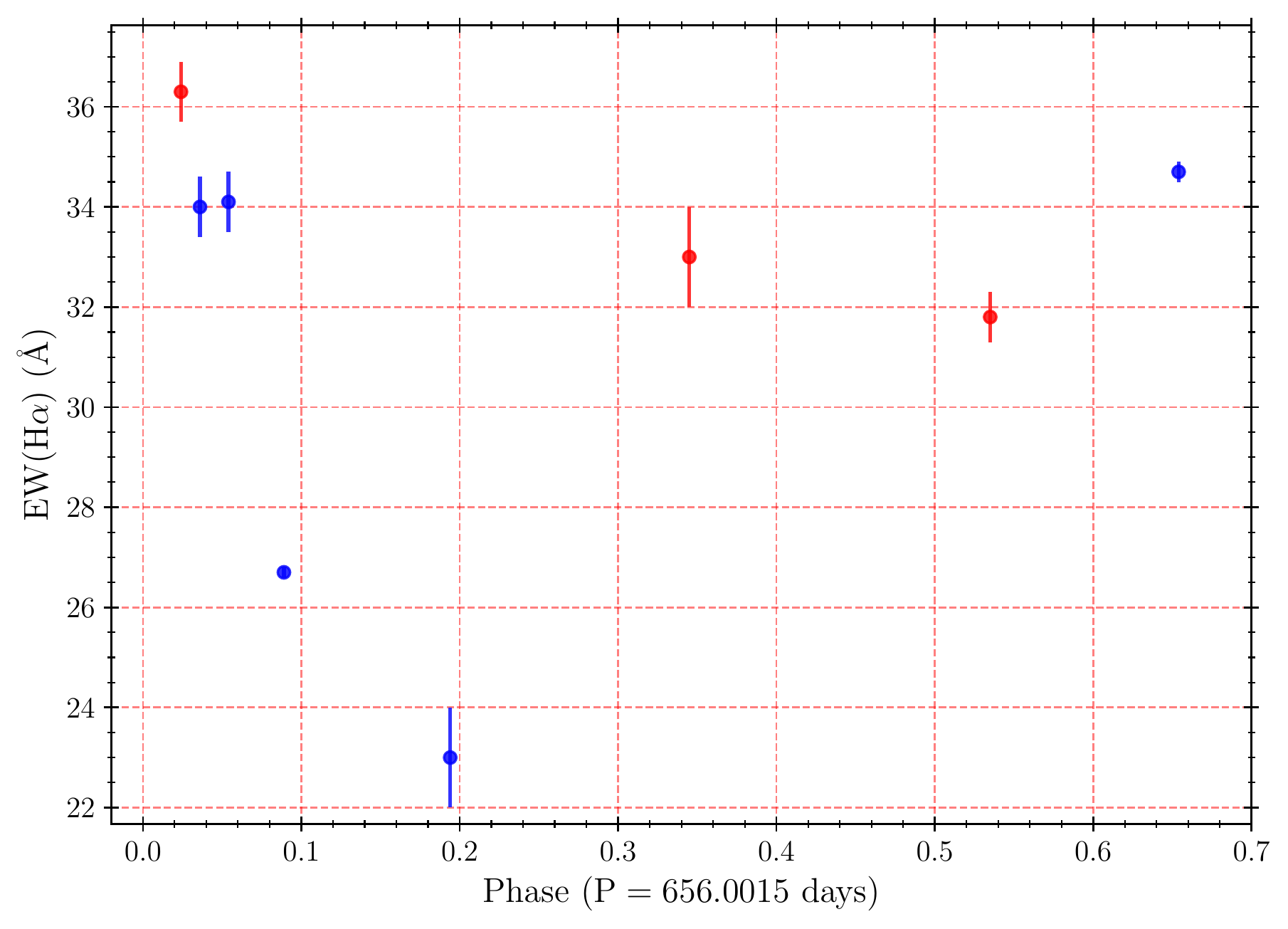}
\end{center}
\caption{Changes in EW(H\,$\alpha$) with the orbital phase. The red data points correspond to 
observations were only spectral regions around the H\,$\alpha$ line were obtained.}
\label{fig:EW}
\end{figure}

Using the obtained orbital ephemerides of SXP\,1062, we plot EW(H\,$\alpha$) as function of the orbital 
phase (see Fig.\,\ref{fig:EW}). One can see that after the periastron passage of the neutron star the EW of
the H\,$\alpha$ line drops by about 40 per cent in $\approx130$\,d and then almost returned to its original 
value in the next $\approx100$\,d. We propose that this behaviour of EW(H\,$\alpha$) reflects the partial 
destruction (or changes in the geometry) of the circumstellar disk caused by the passage of the neutron star 
through it and its restoration in the following few months. It also turned out that the periastron passage 
of the neutron star led to other changes in the spectrum of 2dFS\,3831. In particular, we found that in the 
blue spectrum obtained closest to the periastron epoch (phase 0.036), there is a noticeable emission line of 
He\,{\sc ii} $\lambda$4686 (see the left-hand panel of Fig.\,\ref{fig:star}), which disappeared in the 
spectrum obtained about two weeks later (phase 0.054). We also found that in the spectrum obtained at the 
greatest distance from the periastron (phase 0.654), numerous emission lines of Fe\,{\sc ii} appeared 
(see the right-hand panel of Fig.\,\ref{fig:star}), which were absent in other spectra. We interpret these 
changes in the spectrum as a consequence of the heating of the circumstellar disk due to the passage of the 
neutron star through it and its subsequent cooling.

On a related note, using the four available wide-range spectra, we measured the heliocentric radial 
velocity, $V_{\rm hel}$, of 2dFS\,3831 (see Table\,\ref{tab:rad}). The mean value of these measurements 
of $185\pm5 \, \kms$ is close to the systemic velocity of the SNR shell ($190\pm5 \, \kms$) and that of the 
background H\,$\alpha$ emission ($188\pm6 \, \kms$), which could mean 
that the post-SN binary obtained a low or zero kick velocity along our line-of-sight. At the same time, 
one can see that $V_{\rm hel}$ measured on 2014 June 27 (i.e. soon after the periastron passage) differs 
from the mean value by $4\sigma$. This may be due to some error in the data processing, but we consider this 
unlikely. Our preference is that the noticeable increase in the radial velocity near the periastron is due
to the high eccentricity of the binary orbit. This can be tested by additional radial velocity measurements.

\begin{table}
\centering\caption{Changes in the heliocentric radial velocity of 2dFS\,3831.}
\label{tab:rad}
\begin{tabular}{lcc} \hline
Date             & $V_{\rm hel} (\kms)$ & Phase \\
\hline
2012 October 13  & $184\pm5$  & 0.089 \\
2014 June 27     & $206\pm5$  & 0.036 \\
2014 July 9      & $180\pm5$  & 0.054 \\
2020 December 24 & $170\pm20$ & 0.654 \\ 
\hline
\end{tabular}
\end{table}\

Finally, we note that the maximum value of the absolute value of EW(H\,$\alpha$) (measured for 
2dFS\,3831 in 2014) and the $P_{\rm orb}-$EW(H\,$\alpha$) diagram by Reig et al. (1987) imply that 
the orbital period of SXP\,1062 should be $\approx150$\,d, which is a factor of 4 smaller than the 
observed value. This suggests that the radial size of the disk is not limited by the interaction 
with the neutron star, but by some other factor. We speculate that the equatorial spread of the 
excretion disk might be inhibited by the high thermal pressure in the SNR interior. 

\section{Discussion}
\label{sec:dis}

The measured expansion velocity of the optical shell of $V_{\rm sh}\approx140 \, \kms$ and 
a lace appearance of this shell are typical of SNRs in the radiative (or snow-plough) phase (e.g. 
Lozinskaya 1992). If MCSNR\,J0127$-$7332 is indeed in the radiative phase, then the expansion velocity 
of the SN blast wave $V_{\rm SNR}$ is equal to $V_{\rm sh}$.

Let us compare the observed line ratios listed in Table\,\ref{tab:int} with the theoretical ones 
calculated by Allen et al. (2008) for shock models with the SMC 
abundances\footnote{http://cdsweb.u-strasbg.fr/$\sim$allen/shock.html} 
(see their models started with the letter P). A detailed comparison showed that the observed line ratios 
taken together do not fit any model. Namely, while the velocity sensitive line ratios, such as 
[O\,{\sc i}]/H\,$\alpha$ and [O\,{\sc ii}]/H\,$\beta$, clearly indicate that the shock velocity 
is $\approx100-150 \, \kms$, the high value of the [O\,{\sc iii}]/H\,$\beta$ ratio requires much higher 
shock velocities. Also, although the [S\,{\sc ii}]/H\,$\alpha$ ratio in region\,A could be produced by 
fast ($\geq250 \, \kms$) shocks with high values of the magnetic parameter, the much lower value of this 
ratio measured for region\,B does not fit any of the models. It should be noted, however, that the models 
by Allen et al. (2008) consider the ionizing radiation emitted only by the shocks themselves (i.e. 
autoionizing shocks) and did not account for possible presence of other sources of this radiation, such 
as the donor stars of high-mass X-ray binaries associated with SNRs and/or OB stars in the close 
environments of SNRs.

The contradiction between the SNR expansion velocity estimates based on different line ratios can be 
avoided if some oxygen in the local ISM is doubly ionized. In this case, even a slow shock wave can 
produce strong [O\,{\sc iii}] lines (Raymond 1979). Similarly, the very low [S\,{\sc ii}]/H\,$\alpha$ 
ratio could be understood if a significant part of sulfur in the preshock gas is photoionized to 
S$^{++}$. If MCSNR\,J0127$-$7332 is indeed in a radiative phase, then its slow blast wave does not 
produce a notable photoionized precursor (e.g. Dopita \& Sutherland 1996). Thus, one needs to assume 
that the preshock gas is ionized by radiation from the central star of the SNR and/or OB stars in 
its environs (cf. Section\,\ref{sec:smc}). In this connection, we note that a small value of the 
[S\,{\sc ii}]/H\,$\alpha$ ratio was also found in MCSNR\,J0103$-$7201 (Gvaramadze et al. 2019), which 
along with MCSNR\,J0127$-$7332 are the only two known SNRs in the SMC associated with a BeXB. For 
MCSNR\,J0103$-$7201 we found the expansion velocity of $\sim100 \, \kms$ (Gvaramadze et al. 2019), 
which is too low to explain the presence of the strong [O\,{\sc iii}] $\lambda$5007 emission line in 
the spectrum of this SNR ([O\,{\sc iii}]/H\,$\beta \approx2-3.5$). We speculate that the environs of 
both these SNRs are ionized by their central stars. This possibility, however, has to be proved by 
further research.

Our estimate of the SNR expansion velocity is a factor of three lower than the expansion velocity 
derived by H\'enault-Brunet et al. (2012) and the one following from the X-ray spectral modelling
by Haberl et al. (2012). Let us discuss this discrepancy.

Both H\'enault-Brunet et al. (2012) and Haberl et al. (2012) assumed that MCSNR\,J0127$-$7332 is 
in the Sedov (adiabatic) phase. Apparently, their assumption is based on the widely accepted hypothesis 
by McKee \& Cowie (1975) ``that all the optically observed SNRs are still in the adiabatic phase of 
their expansion'', which was put forward to explain correlation between optical and X-ray emission in 
the SNR Cygnus Loop. This hypothesis suggests that the SN blast wave in Cygnus Loop propagates in a 
cloudy medium, and that the optical emission in this SNR is produced by radiative shocks in dense 
cloudlets, while the X-ray emission originates in the lower-density intercloud medium shocked by the 
adiabatic blast wave (McKee \& Cowie 1975; Bychkov \& Pikelner 1975). Correspondingly, it is assumed 
that the expansion velocity of the SN blast wave, $V_{\rm SNR}$, is related to the temperature of the 
X-ray emitting plasma, $T_{\rm X}$, through the following equation: $V_{\rm SNR}=920 \, \kms 
(T_{\rm X}/1 \, {\rm keV})^{1/2}$. [Note that application of this equation to MCSNR\,J0127$-$7332 with 
its $T_{\rm X}=0.23$\,keV (Haberl et al. 2012) yields $V_{\rm SNR}=440 \, \kms$.] Also, since radial 
velocity measurements indicate (e.g. Minkowski 1958) that the optically-emitting shell in Cygnus Loop 
is expanding at a lower velocity than that inferred from X-ray observations (Tucker 1971), it was 
suggested (McKee \& Cowie 1975) that the shocked cloudlets are accelerated by the SN blast wave to a 
fraction of its velocity. 

Although the above consideration allows to explain the discrepancy between estimates of the SNR 
expansion velocity based on X-ray and optical observations, it faces a problem in explaining the shape 
of the optical filaments, which are actually bumpy thin sheets viewed at different angles (Hester 1987). 
To avoid this problem, McKee \& Cowie (1975) suggested that the cloudlets must have at least one small 
dimension, i.e. they should be in the form of sheets. It remains unclear, however, what the origin of 
these sheet-like cloudlets and why they ``have arranged themselves in such a nicely spherical shell'' 
(McCray \& Snow 1979) such as observed in Cygnus Loop (and some other SNRs, like Vela, S147, etc.). 
The answer suggested by McCray \& Snow (1979; cf. Charles, Kahn \& McKee 1985) is that Cygnus Loop was 
produced by the SN progenitor star and not by the SN itself, meaning that the SN exploded in a cavity,
evacuated by the stellar wind of the SN progenitor (cf. Gvaramadze et al. 2017).

This idea was explored in detail by a number of workers (e.g. Ciotti \& D'Ercole 1988; Tenorio-Tagle et 
al. 1991; Franco et al. 1991), who modelled a SN explosion in a wind-driven bubble created by the SN 
progenitor star. Particularly, it was shown that the evolution of the SN blast wave depends on the mass 
of the shell surrounding the bubble. In the case when the mass of this wind-driven shell (WDS) is more 
than 50 times greater than the mass of the SN ejecta, $M_{\rm ej}$, the SN blast wave merges with the 
WDS and the resulting SNR skips the Sedov phase and enters directly in the radiative phase (e.g. Franco 
et al. 1991), meaning that $V_{\rm SNR}$ is equal to the expansion velocity of the optically-emitting 
shell $V_{\rm sh}$. In this process, the former WDS (now the SNR shell) acquires a kinetic energy of 
$E_{\rm kin}=M_{\rm sh}V_{\rm sh}^2/2 \approx (0.1-0.3)E_{\rm SN}$, where $M_{\rm sh}\approx 
(4\pi/3)R_{\rm sh}^3 m_{\rm H}n_{\rm ISM}$ is the mass of the WDS (i.e. the mass of the ISM gas 
originally contained within a sphere of radius $R_{\rm sh}$), $R_{\rm sh}$ is the radius of the WDS, 
$m_{\rm H}$ is the mass of the hydrogen atom, $n_{\rm ISM}$ is the number density of the local ISM, 
and $E_{\rm SN}$ is the energy of the SN blast wave (Franco et al. 1991). Moreover, the collision of 
the SN blast wave with the WDS leads to the development of the Rayleigh-Taylor instability, resulting 
in dome-like deformations of the shell which, when viewed from different angles, determine the lace 
appearance of some SNRs, such as Vela and S147 (Gvaramadze 1999, 2006). On the other hand, the inner 
layers of the shocked WDSs could be hot enough to produce soft X-ray emission (e.g. Tenorio-Tagle et al. 
1991). 

Based on the above, we propose that MCSNR\,J0127$-$7332 is the result of SN explosion in a cavity 
surrounded by a massive WDS. This proposal provides a natural explanation of the coexistence of the 
slowly expanding optical shell and the soft X-ray emission within it. Assuming that the size of the SNR 
is equal to the size of the WDS, i.e. $R_{\rm SNR}\approx R_{\rm sh}$ (so that the mass of the SNR shell 
is equal to the mass of the pre-existing WDS), and adopting $E_{\rm SN}=10^{51} \, {\rm erg}$ and 
$E_{\rm kin}=(0.1-0.3)E_{\rm SN}$, one finds that $M_{\rm sh}\approx500-1500 \, \msun$ and $n_{\rm 
ISM}\approx0.4-1.2 \, {\rm cm}^{-3}$. Correspondingly, the age of MCSNR\,J0127$-$7332 is approximately 
equal to the crossing time of the wind bubble, i.e. $t_{\rm SNR}\sim R_{\rm SNR}/V_{\rm ej}$, where 
$V_{\rm ej}=(2E_{\rm SN}/M_{\rm ej})^{1/2}$ is the velocity of the SN ejecta of mass $M_{\rm ej}$. To 
estimate $M_{\rm ej}$, we assume that the SN explosion was symmetric (i.e. no natal kick was imparted 
to the new-born neutron star) and that the current orbital eccentricity, $e$, of SXP\,1062 did not change 
much since the SN explosion. In this case, one finds (e.g. Iben \& Tutukov 1997) that 
$M_{\rm ej}=e(M_*+M_{\rm NS})$, where $M_*$ and $M_{\rm NS}$ are the masses of 2dFS\,3831 and its 
companion neutron star, respectively. Adopting $M_*=15\,\msun$ (H\'enault-Brunet et al., 2012) and 
$M_{\rm NS}=1.4\msun$, one finds that $V_{\rm ej}\ga2\,700 \, e^{-1/2} \, \kms$ and $t_{\rm SNR}\la 
8\,000 \, e^{1/2}$\,yr. We caution, however, that these estimates should be regarded only as rough 
order-of-magnitude values.

\section{Summary}
\label{sec:sum}

We have presented the results of optical spectroscopic observations of the SNR MCSNR\,J0127$-$7332 in the 
SMC and the mass donor star, 2dFS\,3831, of its associated BeXB SXP\,1062 carried out with the Southern 
African Large Telescope (SALT). The long-slit SALT spectroscopy of the SNR shell allowed us to measure its 
expansion velocity of $\approx140 \, \kms$, which is typical of SNRs in the radiative (snow-plough) 
phase. A comparison of the line ratios in the spectrum of the SNR shell with the library of line
intensities for shock models (MAPPINGS\,III) showed that the observed line ratios taken together did not 
fit any model. It was suggested that these ratios can be explained if the local ISM is ionized by 2dFS\,3831
and/or by OB stars in the vicinity of the SNR. To reconcile the coexistence of the slowly expanding optical
shell of the SNR and the soft X-ray emission within it, it was proposed that the SN explosion occurred within 
a cavity evacuated by the wind of the SN progenitor star and that the cavity at the moment of SN explosion 
was surrounded by a massive shell (more than 50 times more massive than the SN ejecta). This proposal  
implies the age of the SNR of $\la10,000$\,yr.

Spectroscopic observations of 2dFS 3831 has revealed that EW of the H\,$\alpha$ emission line decreased by 
about 40 per cent in $\approx 130$\,d after the periastron passage of the neutron star and then almost 
returned to its original value in the next $\approx 100$\,d. These changes in EW are accompanied by other 
changes in the spectral appearance of the Be star: in the spectrum obtained shortly after the periastron, 
there was a noticeable emission line of He\,{\sc ii} $\lambda$4686, which disappeared in the next two weeks 
or so. We interpreted these changes as the result of the interaction of the neutron star with the circumstellar 
disk which led to a temporary disturbance and heating of the disk. We also found an indication that the 
neutron star orbits 2dFS 3831 in a highly eccentric orbit, but additional high-resolution spectra covering 
all orbital phases of the binary system are needed to confirm this.

\section{Acknowledgements}
This work is based on observations obtained with the Southern African Large Telescope (SALT), under programs 
\mbox{2012-1-RSA\_UKSC-003}, \mbox{2014-1-RSA\_OTH-022}, 2016-2-SCI-044, 2018-1-MLT-008 and 2020-1-MLT-003. 
VVG acknowledges support from the Russian Science Foundation under grant 19-12-00383. A.Y.K. acknowledges 
support from the National Research Foundation (NRF) of South Africa. LMO acknowledges partial support by the 
Russian Government Program of Competitive Growth of Kazan Federal University. This research has made use of 
the SIMBAD data base, operated at CDS, Strasbourg, France.

\section{Data availability}

The data underlying this article will be shared on reasonable request to the corresponding author.

\end{document}